\documentclass[%
 reprint,
nofootinbib,
 amsmath,amssymb,
 aps,superscriptaddress
]{revtex4-2}

\usepackage{graphicx}
\usepackage{dcolumn}
\usepackage{bm}
\usepackage{mathrsfs}
\usepackage{subfig}

\begin{document}

\title{Features of the Primordial Universe in $f(R)$-gravity as viewed in
the Jordan frame}
\author{Nicola Bamonti}
\affiliation{Physics Department (VEF), Sapienza University of Rome, P.le A. Moro 5 (00185) Roma, Italy}
\author{Andrea Costantini}
\affiliation{Physics Department (VEF), Sapienza University of Rome, P.le A. Moro 5 (00185) Roma, Italy}
\author{Giovanni Montani}
\affiliation{Physics Department (VEF), Sapienza University of Rome, P.le A. Moro 5 (00185) Roma, Italy}
\affiliation{ENEA, Fusion and Nuclear Safety Department,
C.R. Frascati, Via E. Fermi 45 (00044) Frascati (RM), Italy}
\begin{abstract}
We analyze some relevant features of the primordial Universe as viewed
in the Jordan frame formulation of the $f(R)$-gravity, especially
when the potential term of the non-minimally coupled scalar field
is negligible. We start formulating the Hamiltonian picture in the
Jordan frame, using the 3-metric determinant as a basic variable and
we outline that its conjugated momentum appears linearly only in the
scalar constraint. Then, we construct the basic formalism to characterize
the dynamics of a generic inhomogeneous cosmological model and specialize
it in order to describe behaviors of the Bianchi Universes, both on
a classical and a quantum regime. As a fundamental issue, we demonstrate
that, when the potential term of the additional scalar mode is negligible
near enough to the initial singularity, the Bianchi IX cosmology is
no longer affected by the chaotic behavior, typical in vacuum of the
standard Einsteinian dynamics. In fact, the presence of stable Kasner
stability region and its actractive character are properly characterized.
Finally, we investigate the canonical quantization of the Bianchi
I model, using as time variable the non-minimally coupled scalar field
and showing that the existence of a conserved current is outlined for the
corresponding Wheeler-DeWitt equation. The behavior of a localized
wave-packet for the isotropic Universe is also evolved, demonstrating
that the singularity is still present in this revised quantum dynamics. 
\end{abstract}

\maketitle

\part{Introduction}

The Standard Cosmological Model \cite{Kolb-Turner},
which provides a consistent and probed general picture of the Universe
thermal history, at least after a post-inflationary temperature (but
also the inflationary model is a reliable theoretical framework with
observational favourable indications) is entirely based on the standard
Einsteinian formulation of the gravitational field (also the Standard
Model of elementary particles has a fundamental role). However, it
is a well-know fact that, at very primordial instants (say pre-inflationary
age) its nature can be more complicated than the simple isotropic
model and additional effects from modified and quantum
gravity can become important  \cite{montani2014canonical,montani2011primordial,thorne2000gravitation}. The most natural
generalization of the isotropic Robertson geometry is provided by
\cite{landau2013classical, montani2011primordial}. In particular,
the Bianchi type VIII and IX cosmological models (the most general
allowed by the homogeneity constraint) have an Einsteinian chaotic
dynamics, which constitutes a valuable prototype for the asymptotic
behavior to the singularity of the generic inhomogeneous universe
(no space symmetries are imposed) \cite{BKL82, BKL70,montani2011primordial}. Furthermore it must be noted that, while the
kinematic
structure of General Relativity (namely its covariant nature) expressed
via the tensor language is a solid theoretical construction, the
Einstein equations are not very general in their derivation. In fact,
the Einsteinian dynamics is fixed by the simplest scalar action (the
Einstein-Hilbert action), having the special feature of being the
only one to provide field equations with second derivatives of the
metric tensor only. More general formulation for the gravitational
field dynamics can be easily constructed by very general scalar Lagrangian
\cite{RevModPhys.82.451}.
Among the possible restatements of the gravitational dynamics, it
stands out for its simplicity and its capability to explain phenomena
like dark matter and, overall, the so-called $f(R)$ model, whose Lagrangian is a function of the Ricci scalar. The interest for
this model relies also in the possibility to restate the $f(R)$ model
in terms of a scalar-tensor formulation, in which a scalar field is
non-minimally coupled to standard gravity in the so-called \emph{Jordan
frame} or minimally coupled to it in the so-called Einstein frame
 \cite{RevModPhys.82.451,Bamba_2012}.
We finally observe that the canonical quantization of the gravitational
field is also an open problem in the $f(R)$ model, calling attention
as in the case of ordinary Einsteinian gravity. In fact, the idea that
the corrections to the Einstein-Hilbert action must be relevant when
the Universe density overcome the nuclear density and, more in general,
near enough to the initial singularity, must be conjugated with the
idea that, in the same limit, at Planckian scales, the quantum dynamics
of the gravitational field is a mandatory description. In this paper,
we provide a general discussion of the primordial Universe in the
$f(R)$ formulation of the gravitational field, as viewed in the Jordan
frame, facing both symmetric and generic models, as well as both classical
and quantum effects; for related topics see \cite{2013EPJP..128..155C, PhysRevD.90.101503, LecianMontani,  rev_odin}. We start by re-analyzing the study presented
in  \cite{PhysRevLett.106.171301}, in which the Hamiltonian formulation
of the $f(R)$ gravitational dynamics in the Jordan frame is developed.
In particular, we introduce a representation of the 3-metric tensor
in which the 3-determinant is isolated, like in the original analysis
in \cite{PhysRev.160.1113}. Thus, we outline that the conjugate momentum to the 3-determinant enters linearly only in the gravitational Hamiltonian. Considerations
on the structure of the corresponding Wheeler-DeWitt equation are
developed, clarifying how it seems to resembles the morphology of a Schr\"oedinger
functional equation (although non-local) than a Klein-Gordon functional
equation as in standard gravity \cite{montani2014canonical} with respect to the choice
of the 3-determinant as time variable. On the base of the general formulation
above traced, we consider the formulation of a generic cosmological
model in the Jordan frame, generalizing similar studies developed
in Einsteinian gravity, see \cite{PhysRev.117.1595,Kirillov93,montani&imponente, montani&benini2007}. Then, we specialize this scheme
to the symmetric case of the homogeneous but anisotropic Bianchi
universes, for which we construct the generic Hamiltonian picture,
then focusing attention to the Bianchi I and Bianchi IX models. Here
we provide a detailed analysis of the Bianchi IX universe, demonstrating
that, as far as the potential term of the non-minimally coupled scalar
field (depending on the form of $f$ ) is negligible, the chaos of
the standard gravitational dynamics in vacuum is removed. We outline
the existence of the so-called \emph{generalized Kasner solution}
and, via a numerical analysis, we show its actractivity during the
system evolution. Finally, we face the treatment of quantum features
of the $f(R)$ gravity in the Jordan frame, by implementing the Dirac
canonical method to the restated Hamiltonian constraints. We pay attention
on the Bianchi I cosmological model, constructing a conserved current
for the associated Wheeler-DeWitt equation. The construction of an
evolutive wave packet is limited to the isotropic case only, being
that, if the non-minimally coupled scalar field potential is negligible,
the singularity still appears in the generalized quantum gravity framework
of the Jordan frame, having chosen such a field as the system internal
time. The comparison with the case of the standard Einsteinian isotropic
quantum Universe in the presence of a minimally coupled scalar field
as a matter clock is finally provided. \linebreak{}
The manuscript is organized
as follows.
In Part II we present the $f(R)$ theories of gravity in the so-called
\emph{Jordan frame}. We also introduce the $f(R)$ Hamiltonian formalism
and then we quantize it using the Dirac scheme.\linebreak{}
In Part III we provide a general picture of the Bianchi classification, focusing our attention on the Kasner solution for Bianchi I Universe in standard General Relativity and on the Bianchi IX universe. We underline its chaotic behaviour and characterize its evolution as we approaching to cosmological singularity. 
In Part IV we deal with the formalism of the generic cosmological
problem of the $f(R)$ theories in the Jordan frame.\linebreak{}
In Part V we analyze the features of classical $f(R)$ cosmology. We derive Kasner solutions and then we use them to determine the conditions for which the scalar field potential is negligible. We also perform an analysis of the Bianchi IX cosmology showing the attractivity of the stability region. 
In Part VI we perform a critical study on the canonical quantization
of the Bianchi I model and of the homogeneous and isotropic Universe
(FLRW) in the case of the $f(R)$ theories of gravity in the Jordan
frame. In particular, using the non-minimally coupled scalar field
as a quantum time, we carry out an analysis on the quantum dynamics
of the FLRW model and finally we make a comparison with the case of
a minimally coupled external scalar field in General Relativity. In Part VII concluding remarks follow. 

\part{$f(R)$ gravity}

The Einstein's theory of General Relativity represents the current
classical theory of gravity. Its geometrical and tensorial structure
determines the kinematics of the gravitational field in a very consistent
formulation, but its dynamics admits a wide class of different proposals.
In fact the Einstein-Hilbert Lagrangian $\mathcal{L}_{EH}=R$, proposed
by Hilbert in 1915, is only the most simple proposal. The Ricci scalar
$R$ depends on the second derivatives of the metric tensor $g_{\mu\nu}$
and the corresponding field equations constitute a system of second
order differential equations with respect to the metric tensor $g_{\mu\nu}$,
in full coherence with the "aesthetic request"
of the late 19th century to have a physical theory containing at most
second-order derivatives of potentials. Even the choice of a covariant
Lagrangian is not mandatory. In fact, for the equation of motions to
be covariant, the covariance of the action functional is just a sufficient
but not necessary condition.

\section{The Jordan frame}

A more general formulation of the gravitational field dynamics consists
in the replacement of the Ricci scalar by a generic function $f(R)$.
In the $f(R)$ theories of gravity, proposed by Buchdal in 1970, the
corresponding action of the gravitational field takes the form 
\begin{equation}
S_{g}=\cfrac{1}{2k}\int d^{4}x\sqrt{-g}\,f(R),
\label{1.1}
\end{equation}
where $k=8\pi Gc^{-4}$ and $-g$ is the positive determinant of the
metric tensor $g_{\mu\nu}$. In this way, a new geometric degree of
freedom is included in the theory (the function $f$), which induces
different field equations \cite{RevModPhys.82.451}. 
There are actually three possible variational principles that one
can apply in order to derive the Einstein's field equations . In the
following, we will refer exclusively to the metric formalism. Beginning
from the action \eqref{1.1} and adding a matter term $S_{M}$, the total
action for $f(R)$ gravity takes the form 
\begin{equation}
S=\cfrac{1}{2k}\int d^{4}x\sqrt{-g}f(R)+S_{M}\left(g_{\mu\nu},\psi\right),
\label{1.2}
\end{equation}
where $\psi$ denotes a generic matter field. 

In a field theory
it is always possible to perform redefinitions of the fields, in order
to rewrite the action and the equations in a more convenient way.
If the new action coincides with the previous one, the two theories
are said to be dynamically equivalent and they are actually just different
representations of the same theory. In what follows, we will show
the equivalence between the metric $f(R)$ gravity theory and a specific
theory within the Brans-Dicke class. One can introduce an auxiliary
field $A$ and rewrite the action \eqref{1.2} as 
\begin{equation}
S=\cfrac{1}{2k}\int d^{4}x\sqrt{-g}\left[f(A)+f'(A)(R-A)\right]+S_{M}.
\label{1.4}
\end{equation}
Variation with respect to $A$ leads to the equation 
\begin{equation}
f''(A)\left(R-A\right)=0.
\label{1.5}
\end{equation}
Therefore, if $f''(A)\neq0$ we obtain the result $A=R$ and by substitution
in the action \eqref{1.4} we find the dinamically equivalent action \eqref{1.2}.
We now redefine the field $A$ by introducing the scalar field $\phi=f'(A)$
and its potential $V\left(\phi\right)=A(\phi)\phi-f\left(A(\phi)\right)$.
In this way the action \eqref{1.4} takes the form 
\begin{equation}
S=\cfrac{1}{2k}\int d^{4}x\sqrt{-g}\left[\phi R-V(\phi)\right]+S_{M}\left(g_{\mu\nu},\psi\right).
\label{1.6}
\end{equation}
This is the Jordan frame representation of the action of a Brans--Dicke
theory with Brans--Dicke parameter $\omega_{0}=0$. Action \eqref{1.6}
is also known as the $f(R)$ gravitational action in the Jordan frame.
It is the scalar-tensor equivalent representation of the action \eqref{1.2},
expressed through a non-minimal coupling between a scalar field $\left(\phi\right)$
and the curvature $\left(R\right)$.

\section{Jordan frame Hamiltonian of $f(R)$ gravity}

In this section we will review the Arnowitt-Deser-Misner Hamiltonian formulation of
$f(R)$ gravity in the Jordan frame using the embedding technique.
Consider a four dimension spacetime $\mathcal{M}^{4}$ diffeomorphic
to a manifold $\mathbb{R}\text{\texttimes}\Sigma^{3}$, allowing the
splitting $\mathcal{M}^{4}=\mathbb{R}\text{\texttimes}\Sigma^{3}$.
In this way $\mathcal{M}^{4}$ can be foliated by a one-parameter
family of space-like 3-hypersurfaces $\Sigma_{t}^{3}$, embedded in
the ambient spacetime and defined by mean a temporal scalar parameter
$t$. We introduce the extrinsic curvature 
\begin{equation}
K_{ij}=\frac{1}{2N}\left(D_{i}N_{j}+D_{j}N_{i}-\partial_{t}h_{ij}\right),
\label{2.1}
\end{equation}
where $D_{i}$ ($i=1,2,3$ spatial indices) is the covariant derivative
associated to the 3-metric $h_{ij}$ induced on each $\varSigma_{t}$.
The Ricci scalar can be written in terms of the ADM variables $\left(N,N^{i},h_{ij},t\right)$
through the so-called Gauss-Codazzi equation 
\begin{align}
R & =\,_{T}K.\,_{T}K-\frac{2}{3}K^{2}+\nonumber \\
 & \quad+\,^{3}R+\frac{2}{\sqrt{-g}}\partial_{\mu}\left(\sqrt{-g}n^{\mu}K\right)\nonumber \\
 & \quad-\frac{2}{N\sqrt{h}}\partial_{i}\left(\sqrt{h}h^{ij}\partial_{j}N\right),
\label{2.2}
\end{align}
where $^{3}R\left(h_{ij},\partial_{l}h_{ij},\partial_{m}\partial_{l}h_{ij}\right)$
is the Ricci 3-scalar, $\,_{T}K_{ij}=\left(K_{ij}-\frac{1}{3}h_{ij}K\right)$
; $\,_{T}K.\,_{T}K=\,_{T}K_{ij}\,_{T}K^{ij}$ and $h=det\left(h_{ij}\right)$.
Substituting the relation \eqref{2.2} in the total Action \eqref{1.6} and neglecting
the matter term $S_{M}$, we obtain the ADM action of $f(R)$ gravity
in the Jordan frame (with $k\equiv1)$ 
\begin{equation}
\begin{array}{l}
S_{JF}^{ADM}\left(h_{ij},N,N^{i},\phi\right)=\\
\int_{\mathbb{R}\text{\texttimes}\Sigma}dtd^{3}xN\sqrt{h}\Biggl[\cfrac{\phi}{2}\left(\,_{T}K.\,_{T}K-\frac{2}{3}K^{2}+\,^{3}R\right)-\cfrac{1}{2}V\left(\phi\right)\\
-\cfrac{K}{N}\left(\dot{\phi}-N^{i}\partial_{i}\phi\right)+\frac{1}{N}\partial_{i}\phi\partial^{i}N\Biggr].
\end{array}
\label{2.3}
\end{equation}
Consequently, we immediately write down the $f(R)$ Lagrangian density
in terms of the ADM variables 
\begin{equation}
\begin{array}{l}
\mathcal{L}_{JF}^{ADM}\left(h_{ij},N,N^{i},\phi\right)=\\
N\sqrt{h}\Biggl[\cfrac{\phi}{2}\left(\,_{T}K.\,_{T}K-\frac{2}{3}K^{2}+\,^{3}R\right)-\cfrac{1}{2}V\left(\phi\right)\\
-\cfrac{K}{N}\left(\dot{\phi}-N^{i}\partial_{i}\phi\right)+\frac{1}{N}\partial_{i}\phi\partial^{i}N\Biggr].
\end{array}
\label{2.4}
\end{equation}
Conjugated momenta to the dynamical variables $h_{ij}$ and $\phi$
are defined as 
\begin{equation}
\begin{array}{ccl}
p^{ij} & = & \cfrac{\partial\mathcal{L}_{JF}^{ADM}}{\partial\dot{h}_{ij}}\\
 & = & \cfrac{\sqrt{h}}{2}\left[\phi\left(\,_{T}K^{ij}-\frac{2}{3}Kh^{ij}\right)-\cfrac{h^{ij}}{N}\left(\dot{\phi}-N^{k}\partial_{k}\phi\right)\right]\\
\\
\pi_{\phi} & = & \cfrac{\partial\mathcal{L}_{JF}^{ADM}}{\partial\dot{\phi}}=-\sqrt{h}K.
\end{array}
\label{2.5}
\end{equation}
Expressing the generalized velocities $\left\{ \dot{h}_{ij},\dot{\phi}\right\} $
in terms of the canonical variables 
\begin{equation}
\dot{h}_{ij}=\cfrac{N}{\sqrt{h}}\left(\cfrac{4\,_{T}p_{ij}}{\phi}-\cfrac{2}{3}h_{ij}\pi_{\phi}\right)+D_{i}N_{j}+D_{j}N_{i};
\label{2.6}
\end{equation}
\begin{equation}
\dot{\phi}=\cfrac{2N}{3\sqrt{h}}\left(\phi\pi_{\phi}-p\right)+N^{k}\partial_{k}\phi,
\label{2.7}
\end{equation}
the canonical Lagrangian density is therefore 
\begin{equation}
\begin{array}{l}
\mathcal{L}_{JF}^{ADM}\left(p^{ij},h_{ij},\pi_{\phi},\phi\right)=\\
N\sqrt{h}\Biggl[\cfrac{2}{h\phi}\,_{T}p.\,_{T}p-\cfrac{\phi\pi_{\phi}^{2}}{3h}+\cfrac{\phi\,^{3}R}{2}-\cfrac{V(\phi)}{2}\\
+\cfrac{2\pi_{\phi}\left(\phi\pi_{\phi}-p\right)}{3h}+\cfrac{1}{N}\left(\partial_{i}\phi\partial^{i}N\right)\Biggr],
\end{array}
\label{2.8}
\end{equation}
where $p=h_{ij}p^{ij}$ and $\,_{T}p.\,_{T}p=\,_{T}p_{ij}\,_{T}p^{ij}$.
In analogy with the articles \cite{PhysRevLett.106.171301, PhysRevD.80.084032, Bombacigno:2019nua}
, by performing a Legendre transformation, we obtain the following
Hamiltonian density 
\begin{equation}
\begin{array}{l}
\mathcal{H}_{JF}^{ADM}=N\Biggl[\cfrac{2}{\sqrt{h}}\left(\cfrac{\,_{T}p.\,_{T}p}{\phi}+\cfrac{1}{6}\phi\pi_{\phi}^{2}-\cfrac{1}{3}p\pi_{\phi}\right)\\
+\cfrac{\sqrt{h}}{2}\left(V(\phi)-\phi\,^{3}R+2D_{i}D^{i}\phi\right)\Biggr]+N^{i}\left[\pi_{\phi}\partial_{i}\phi-2D^{j}p_{ij}\right]\\
=N\mathcal{H}_{g,JF}+N^{i}\mathcal{H}_{i,JF}^{g},
\end{array}
\label{2.9}
\end{equation}
where we have highlighted respectively the so-called \emph{superHamiltonian}
$\mathcal{H}_{g,JF}$ and \emph{supermomentum} $\mathcal{H}_{i,JF}^{g}$
which represent the two secondary constraints of the theory as we
can see from the Hamilton's equations 
\begin{equation}
\left\{ \begin{array}{cc}
\delta_{N}S_{JF}^{ADM}=0\rightarrow-\cfrac{\partial\mathcal{H}_{JF}^{ADM}}{\partial N}=0 & \longrightarrow\mathcal{H}_{g,JF}=0\\
\\
\delta_{N^{i}}S_{JF}^{ADM}=0\rightarrow-\cfrac{\partial\mathcal{H}_{JF}^{ADM}}{\partial N^{i}}=0 & \longrightarrow\mathcal{H}_{i,JF}^{g}=0
\end{array}\right.
\label{2.10}
\end{equation}
The prefix \emph{super}, conceived by Wheeler, indicates that the
configurational space of canonical gravity is the space of all Riemannian
3-metrics $Riem\left(\varSigma\right)$ modulo the spatial diffeomorphisms
group $Diff\left(\varSigma\right)$ on the slicing surface $\varSigma$.
Explicitly 
\begin{equation}
\left\{ h_{ij}\right\} =\cfrac{Riem\text{(\ensuremath{\Sigma)}}}{Diff(\Sigma)}.
\label{2.11}
\end{equation}
This is the space of all 3-geometries and it is known as the \emph{Wheeler
superspace}: the $\infty-$dimensional functional space of all the
equivalence classes of 3-metrics $h_{ij}\left(t,x^{l}\right)$ linked
together by 3-diffeomorphisms.

\section{Wheeler-De Witt equation}

The canonical theory of gravity, also in its $f(R)$ extension, can
be written as a dynamical system subjected to first-class constraints
with a Dirac algebra. To implement the quantization of such constrained
system we follow the Dirac scheme, which consists in imposing the
first-class constraints of the theory as quantum operators that annihilate
physical states. 
\begin{equation}
\hat{\mathcal{C\,}}|\Psi\rangle=0.
\label{3.1}
\end{equation}
Proceeding in a more formal way, we impose the constraints \eqref{2.10}
as quantum operators 
\begin{equation}
\begin{array}{c}
\mathcal{\hat{H}}_{g,JF}|\Psi\rangle=0,\\
\mathcal{\hat{H}}_{i,JF}^{g}|\Psi\rangle=0,
\end{array}
\label{3.2}
\end{equation}
where $\Psi$ is known as the wave function of the Universe. More
precisely, in \emph{Wheeler's superspace} the physical states are
wave functionals 
\begin{equation}
\Psi\left[\left\{ h_{ij}\right\} ,\phi\right].
\label{3.3}
\end{equation}

In \emph{Wheeler's superspace}, the dynamics of the gravitational
field is generated by imposing the \emph{superHamiltonian} constraint,
which leads to the Wheeler-De Witt (WDW) equation. It is well known
that in General Relativity (GR) the WDW equation can be seen as a
Klein-Gordon (KG) equation with a varying mass. By defyining the new
set of variables $\left\{ \xi,u_{ij}\right\} $ 
\begin{equation}
\Biggl\{\begin{array}{l}
\xi=h^{\frac{1}{4}}\\
h_{ij}=\xi^{\frac{4}{3}}u_{ij},u=det(u_{ij})=1
\end{array}
\label{3.7}
\end{equation}
and imposing the canonical transformation 
\begin{equation}
\pi_{\xi}\partial_{t}\xi+\pi^{ij}\partial_{t}u_{ij}=p^{ij}\partial_{t}h_{ij},
\label{3.8}
\end{equation}
we can immediately obtain the WDW-KG-like equation in GR 
\begin{align}
\mathcal{\hat{H}}_{g}|\Psi\rangle & =\left[\cfrac{2k}{\xi^{2}}\left(\hat{\pi}^{ij}\hat{\pi}^{lm}u_{il}u_{jm}-\cfrac{3}{32}\xi^{2}\hat{\pi}_{\xi}^{2}\right)-\cfrac{\sqrt{h}}{2k}\,^{3}R\right]|\Psi\rangle\nonumber \\
 & =2k\hbar^{2}\Biggl[\cfrac{3}{32}\cfrac{\delta^{2}}{\delta\xi^{2}}-\cfrac{1}{\xi^{2}}u_{il}u_{jm}\cfrac{\delta^{2}}{\delta u_{ij}\delta u_{lm}}+\nonumber \\
 & -\cfrac{\xi^{2}}{\left(2k\hbar\right)^{2}}\,^{3}R\Biggr]\Psi\left(\xi,\left\{ u_{ij}\right\} \right)=0,
 \label{3.9}
\end{align}
where the variable $\xi$ is timelike (internal time), while the variables
$u_{ij}$ are spacelike. We
now review the corresponding formalism in the case of the $f(R)$
theory in the Jordan frame. Following the same procedure described
above, the \emph{superhamiltonian} $\mathcal{H}_{g,JF}$ in the new
canonical variables $\left\{ \xi,\pi_{\xi};u_{ij},\pi_{ij};\phi,\pi_{\phi}\right\} $
takes the form 
\begin{align}
\mathcal{H}_{g,JF} & =\cfrac{2}{\xi^{2}}\Biggl[\cfrac{1}{\phi}\left(\pi^{ij}\pi^{lm}u_{il}u_{jm}\right)+\cfrac{1}{6}\phi\pi_{\phi}^{2}-\cfrac{1}{4}\xi\pi_{\xi}\pi_{\phi}\nonumber \\
 & +\cfrac{\xi^{4}}{4}\left(V(\phi)-\phi\,^{3}R\left(\xi,u_{ij},\partial\xi,\partial u_{ij}\right)+2D_{i}D^{i}\phi\right)\Biggr].
\label{3.10}
\end{align}
Applying the Dirac quantization, we obtain the WDW-KG-like equation
in the Jordan frame 
\[
\begin{array}{cl}
\mathcal{\hat{H}}_{g,JF}|\Psi\rangle & =\cfrac{2}{\xi^{2}}\Biggl[\cfrac{1}{\phi}\left(\hat{\pi}^{ij}\hat{\pi}^{lm}u_{il}u_{jm}\right)+\cfrac{1}{6}\phi\hat{\pi}_{\phi}^{2}-\cfrac{1}{4}\xi\hat{\pi}_{\xi}\hat{\pi}_{\phi}+\\
 & +\cfrac{\xi^{4}}{4}\left(V(\phi)-\phi\,^{3}R+2D_{i}D^{i}\phi\right)\Biggr]|\Psi\rangle,
\end{array}
\]
\begin{equation}
\begin{array}{cc}
0= & 2\hbar^{2}\Biggl[-\cfrac{1}{\xi^{2}\phi}u_{il}u_{jm}\cfrac{\delta^{2}}{\delta u_{ij}\delta u_{lm}}-\cfrac{\phi}{6\xi^{2}}\cfrac{\delta^{2}}{\delta\phi^{2}}+\cfrac{1}{4\xi}\cfrac{\delta^{2}}{\delta\xi\delta\phi}+\\
 & +\cfrac{\xi^{2}}{4\hbar^{2}}\left(V(\phi)-\phi\,^{3}R+2D_{i}D^{i}\phi\right)\Biggr]\Psi\left(\xi,\left\{ u_{ij}\right\} ,\phi\right).
\end{array}
\label{3.11}
\end{equation}
We can immediately notice that we cannot identify a D'Alembert operator
in the equation, so it is not straightforward to identify a phase
space variable playing the role of a quantum time in the theory. One
result that deserves to be taken seriously is the linearity of the
equation with respect to the conjugate momentum $\pi_{\xi}$. Differently
from the case of General Relativity, the equation \eqref{3.11} is linear
in $\pi_{\xi}$. This makes the Wheeler-De Witt equation in the Jordan
frame conceptually more similar to a Schr\"odinger-like equation, in
which $\xi$ would be the time-like variable. However, there are substantial
differences between the equation \eqref{3.11} and a Schrodinger one, such
as the functional nature and the undefined role of variables $\left\{ \xi,\phi\right\} $
as spacelike or timelike variables of the theory. In fact, there is
a non-trivial coupling between the variables $\left\{ \xi,\phi\right\} $,
which is expressed in the third term containing mixed partial derivatives.
It is also evident that, even if we choose $\xi$ or $\phi$ as a
quantum timelike variable, we will have to face the impossibility
of performing a frequency separation procedure. This critical feature
is due to the presence of the potential term $V(\phi)-\phi\,^{3}R+2D_{i}D^{i}\phi$
which depends on $\left\{ \xi,\phi\right\} $ and their derivatives.

\part{ Bianchi Models}

The Bianchi models are a class of cosmologies that obey to the homogeneity
constraint but not to the isotropy one; they are a generalization
of the FLRW model in which the three independent spatial direction
expand with different rates, introducing a degree of anisotropy. The
relevance of the dynamics of Bianchi models consists in the role these
geometries could have played in a very primordial Universe, \emph{i.e.}
before the inflation phase. 
The anisotropy of the Bianchi models is not only dynamical, due to
the different scaling of different directions, but it has an intrinsic
geometrical nature. We will focus our attention on the Bianchi IX model because
we are looking  for the most general cosmology allowed by the homogeneity
constraint, since we do not have any clue to the need of specific symmetries
and because it also admits the isotropic limit.

\section{ General behaviour of Bianchi models in Misner variables}

We start with the line element of a general Bianchi model\cite{montani2014canonical}: 
\begin{equation}
ds^{2}=-N^{2}(t)dt^{2}+a^{2}(t)(\omega^{l})^{2}+b^{2}(t)(\omega^{m})^{2}+c^{2}(t)(\omega^{n})^{2},
\label{4.1}
\end{equation}
where the 1-forms $\omega^{a}$ obey to the condition 
\begin{equation}
\partial_{[i}\omega_{j]}^{a}=-\frac{1}{2}C_{bc}^{a}\omega_{i}^{b}\omega_{j}^{c}.
\label{4.2}
\end{equation}
The 1-forms $\omega^{a}=\omega_{i}^{a}dx^{i}$ ($a=l,m,n$) fix the
particular geometry of the considered Bianchi model.
The Jacobi identity associated
with the structure constants defines the Bianchi classification of
the models. In order to analyze the general behaviour of the models
we will perform a variables change, rewriting the expansions factor
$a(t),b(t)$ and $c(t)$ using the \emph{Misner variables}: 
\begin{equation}
\begin{split} & \ln a=\alpha+\beta_{+}+\sqrt{3}\beta_{-},\\
 & \ln b=\alpha+\beta_{+}-\sqrt{3}\beta_{-},\\
 & \ln c=\alpha-2\beta_{+}.
\end{split}
\label{4.3}
\end{equation}
The choice of these variables allows us to separate the isotropic contribution,
related to the variables $\alpha$ \emph{i.e.} the logarithm of the
Universe's volume, from the two degrees of anisotropy related to $\beta_{\pm}$.
The \emph{Misner variables} make the kinetic part of the Hamiltonian
diagonal \cite{thorne2000gravitation}. Adopting the coordinate $\alpha,\beta_{\pm}$ the action
takes the form: 
\begin{equation}
S_{B}=\int dt\{p_{\alpha}\frac{d\alpha}{dt}+p_{+}\frac{d\beta_{+}}{dt}+p_{-}\frac{d\beta_{-}}{dt}-cNe^{-3\alpha}\mathcal{H}_B\},
\label{4.4}
\end{equation}
where $c$ is a quantity depending on fundamental constants and on
the particular space integral for the considered type, $p$ and $p_{\pm}$
are the conjugate momenta to $\alpha,\beta_{\pm}$ respectively. So
the super-Hamiltonian takes the form:

\begin{equation}
\mathcal{H}_{B}=-p_{\alpha}^{2}+p_{+}^{2}+p_{-}^{2}+e^{4\alpha}V_{B}\bigl(\beta_{\pm}\bigl).
\label{4.5}
\end{equation}
Where $V_{B}$ denotes a potential term, different for each Bianchi model
and due to the spatial curvature and $\alpha$ seems to be the right
time-like variable of the system. Variating the action with respect
to the lapse-function N, we get the Hamiltonian constraint $\mathcal{H}=0$.
 The Hamiltonian constraint can
be solved in this way\cite{PhysRevLett.22.1071}: 
\begin{equation}
p_{\alpha}=-h_{\alpha}=-\sqrt{p_{+}^{2}+p_{-}^{2}+e^{4\alpha}V_B(\beta_{\pm})},
\label{4.6}
\end{equation}
so the action of a Bianchi model can be rewritten as follows 
\begin{equation}
S_{ADM}=\int d\alpha\{p_{+}\beta_{+}^{'}+p_{-}\beta_{-}^{'}-h_{\alpha}\},
\label{4.7}
\end{equation}
where $\beta_{\pm}^{'}=\frac{d\beta_{\pm}}{d\alpha}$. By neglecting
the potential term of the ADM-reduction of the Hamiltonian we can
find a solution for the Bianchi I model, called \emph{Kasner solution}.
Using the Hamilton's equations we can find the behaviour of $\beta_{\pm}$
as functions of $\alpha$ 
\begin{equation}
\frac{d\beta_{\pm}}{d\alpha}=\frac{p_{\pm}}{p_{\alpha}}=const,
\label{4.8}
\end{equation}
with $|d\beta_{\pm}/d\alpha|^{2}=1$.

\section{ Bianchi IX cosmology}

The Bianchi IX model is the most general cosmology allowed by the
homogeneity constraint and it corresponds to dealing with all three
structure constants different from zero. The Bianchi type IX is associated
to a physical space which is invariant under the $SO(3)$ group of
motion and the 1-forms that defines the geometry takes the following
form\cite{thorne2000gravitation}: 
\begin{equation}
\begin{split} & \omega^{1}=\cos\psi d\theta+\sin\psi\sin\theta d\phi\\
 & \omega^{2}=\sin\psi d\theta-\cos\psi\sin\theta d\phi\\
 & \omega^{3}=d\psi+\cos\theta d\phi.
\end{split}
\label{5.1}
\end{equation}
By redefining the origin of the variable $\alpha$ as $\alpha\rightarrow\alpha-(1/4)\ln(6\pi)$
the constant $c$ is fixed to $\sqrt{3\pi/8}$ and the potential
term takes the following form: 
\begin{equation}
\begin{split}2 & V_{IX}=\left(e^{4\alpha+4\beta_{+}+4\sqrt{3}\beta_{-}}+e^{4\alpha+4\beta_{+}-4\sqrt{3}\beta_{-}}+e^{4\alpha-8\beta_{+}}\right)\\
 & -2\left(e^{4\alpha+4\beta_{+}}+e^{4\alpha-3\beta_{+}+2\sqrt{3}\beta_{-}}+e^{4\alpha-2\beta_{+}-2\sqrt{3}\beta_{-}}\right).
\end{split}
\label{5.2}
\end{equation}
The first parenthesis dominates as we are going to the singularity ($\alpha\rightarrow-\infty$) and the potential term is reduced
to an infinite well with the form of an equilateral curvilinear triangle.
One of the three equivalent side of the triangle is described by the
asymptotic form 
\begin{equation}
V_{IX}\sim\frac{1}{8}e^{4\alpha-8\beta_{+}},\quad\beta_{+}\rightarrow\infty.
\label{5.3}
\end{equation}
Clearly the potential wall moves out as $\alpha$ goes forward, leaving
a potential-free region proportional to $\alpha^{2}$. To determine the velocity of the walls we can simply study the behaviour of the third wall and deduce the behaviour of the others by the symmetries of the equilateral triangle. 

\begin{figure}[h]
\begin{centering}
\includegraphics[scale=0.3]{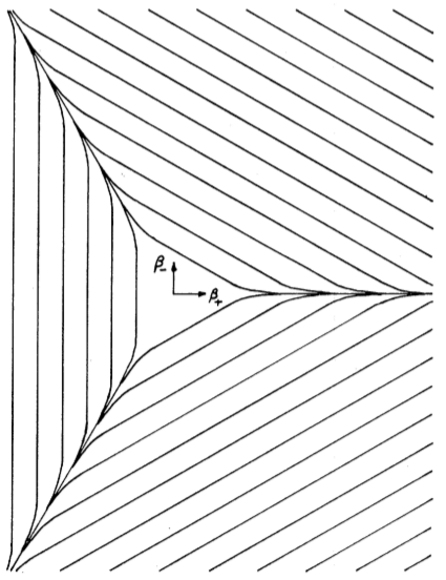} 
\par\end{centering}
\caption{\label{fig1}Equipotential lines of Bianchi IX model in $(\beta_+,\beta_-)$ plane \cite{PhysRev.186.1319}}

\end{figure}

By imposing that the third wall is equal to 1 we can deduce
\begin{equation}
-8\beta_{+}^{wall}+4\alpha=0 \Longrightarrow \beta_+^{wall}=\frac{1}{2}\alpha,
\label{5.4} 
\end{equation}
where $\beta^{wall}$ defines an equipotential in the $\beta$ plane bounding
the region in which the potential (space curvature) term is significant.
The potential wall moves outward with the speed 
\begin{equation}
\biggl|\frac{\partial\beta_{wall}}{\partial\alpha}\biggr|=\frac{1}{2},
\label{5.5}
\end{equation}
the velocity of the point-Universe is equal to one according to \eqref{4.8}
in free-potential region and so is doomed to impact against the wall.
This means that the vacuum Bianchi IX classical evolution is constituted
by an infinite series of Kasner regimes (free motions of the point-Universe),
associated with continuous scatterings against the infinite potential
walls\cite{PhysRev.186.1319}.

We need more details on the "bounce" and we can derive them from
the ADM-Hamiltonian using the asymptotic form of the potential: 
\begin{equation}
\mathcal{H}_{IX}=\sqrt{p_{+}^{2}+p_{-}^{2}+1/3\exp{-4\alpha-8\beta_{+}}},
\label{5.6}
\end{equation}
showing that $\mathcal{H}$ is independent from $\beta_{-}$ so $p_{-}$ is a constant
of motion. Another constant can be found comparing the following equation
\begin{equation}
\begin{split} & p_{+}'=-\partial \mathcal{H}_{IX}/\partial\beta_{+}=+4(3\mathcal{H}_{IX})^{-1}\exp{-4\alpha-8\beta_{+}},\\
 & \mathcal{H}^{'}_{IX}=-\partial \mathcal{H}_{IX}/\partial\alpha=-2(3\mathcal{H}_{IX})^{-1}\exp{-4\alpha-8\beta_{+}},
\end{split}
\label{5.7}
\end{equation}
so we can define $k=1/2p_{+}+\mathcal{H}_{IX}=const$. These two constant of motion
allow us to find the value of $\beta_{+}'$ and $\beta_{-}'$ after
the bounce as function of their value before. Since $|\beta'|=1$
we can parametrize $\beta_{-}'=\sin{\theta}$ and $\beta_{+}'=\cos{\theta}$.
Using \eqref{4.8} and the two constant of motion we get  
\begin{equation}
\begin{split}\mathcal{H}^{i}_{IX}\sin{\theta_{i}}= & \mathcal{H}^{f}_{IX}\sin{\theta_{f}},\\
\mathcal{H}^{i}_{IX}\bigl(-1/2\cos{\theta_{i}}+1\bigr)= & \mathcal{H}^{f}_{IX}\bigl(+1/2\cos{\theta_{f}}+1\bigr).
\end{split}
\label{5.8}
\end{equation}
These can be combined to find an equation for $\theta_{f}$ as function
of $\theta_{i}$, 
\begin{equation}
\sin{\theta_{f}}-\sin{\theta_{i}}=\frac{1}{2}\sin{\left( \theta_{f}+\theta_{i} \right)},
\label{5.9}
\end{equation}
which is sufficient for our purpose.

\part{The generic cosmological problem in the Jordan Frame}

The generic cosmological problem represents the Hamiltonian formalism
of gravity for an inhomogeneous space. The first studies date back
to the '60s when the Landau school started to investigate the properties
and the behaviour of the generic cosmological solution of the Einstein
field equations.

\section{Hamiltonian formulation in a general framework}

Using the vierbein formalism, the 3-metric associated to a generic
inhomogeneus space can be written as 
\begin{equation}
h_{ij}(t,x^{k})=\eta_{ab}(t,x^{k})e_{i}^{a}(t,x^{k})e_{j}^{b}(t,x^{k}),
\label{6.1}
\end{equation}
where $\eta_{ab}$ is its triadic representation and $a=1,2,3$ are
the Lorentz indices. The corresponding line element is 
\begin{align}
ds^{2} & =-N(t,x^{k})dt^{2}\nonumber \\
 & \quad+h_{ij}(t,x^{k})\left(dx^{i}+N^{i}(t,x^{k})dt\right)\left(dx^{j}+N^{j}(t,x^{k})dt\right).
 \label{6.2}
\end{align}
We can define a generic set of 3-vectors on the hypersurfaces $\varSigma_{t}^{3}$
of the ADM foliation as 
\begin{equation}
e_{i}^{a}(t,x^{k})=O_{b}^{a}(x^{k})\partial_{i}y^{b}(t,x^{k}),
\label{6.3}
\end{equation}
where $y^{b}$ denotes three scalar functions and $O_{b}^{a}$ is
a $SO(3)$ matrix such that $O_{b}^{a}O_{a}^{c}=\delta_{b}^{c}$.
This definition of the 3-vectors $e_{i}^{a}$ allows us to rewrite
the 3-metric tensor $h_{ij}$ as 
\begin{align}
h_{ij}(t,x^{k}) & =\underset{a}{\sum}e^{q_{a}(t,x^{k})}e_{i}^{a}(t,x^{k})e_{j}^{a}(t,x^{k})\nonumber \\
 & =\underset{a}{\sum}e^{q_{a}(t,x^{k})}O_{b}^{a}(x^{k})\partial_{i}y^{b}(t,x^{k})O_{c}^{a}(x^{k})\partial_{i}y^{c}(t,x^{k}),
\label{6.4}
\end{align}
where $q_{a}(t,x^{k})=\left\{\ln{(a^{2}(t,x^{k}))},\ln{(b^{2}(t,x^{k}))},\ln{(c^{2}(t,x^{k}))}\right\}$
denotes the three inhomogeneus cosmological scale factors. Imposing
the canonical transformation $\left(h_{ij},p^{ij}\right)\longrightarrow\left(q_{a},p_{a};y_{a},\pi_{a}\right)$
\begin{equation}
p^{ij}\partial_{t}h_{ij}=\underset{a}{\sum}p_{a}\partial_{t}q_{a}\,+\,\underset{a}{\sum}\pi_{a}\partial_{t}y_{a},
\label{6.5}
\end{equation}
we obtain the \emph{superHamiltonian} $\mathscr{H}_{g,JF}$ and the
\emph{supermomentum $\mathscr{H}_{i,JF}^{g}$} in the new generic
framework (see Appendix A for details) 
\begin{align}
\mathscr{H}_{g,JF} & =\cfrac{2}{\sqrt{h}}\left(\cfrac{\underset{a}{\sum}p_{a}^{2}-\frac{1}{3}\left(\underset{a}{\sum}p_{a}\right)^{2}}{\phi}+\cfrac{1}{6}\phi\pi_{\phi}^{2}-\cfrac{1}{3}\underset{a}{\sum}p_{a}\pi_{\phi}\right)\nonumber \\
 & +\cfrac{\sqrt{h}}{2}\left(V(\phi)-\phi\,^{3}R+2D_{i}D^{i}\phi\right),
\label{6.6}
\end{align}
\begin{equation}
\mathscr{H}_{i,JF}^{g}=\pi_{\phi}\partial_{i}\phi+\pi_{a}\partial_{i}y^{a}+p_{a}\partial_{i}q^{a}+2p_{a}\left(O^{-1}\right)_{a}^{b}\partial_{i}O_{b}^{a}.
\label{6.7}
\end{equation}

\section{The generic cosmological problem in Misner-like variables}

Let us introduce the Misner-like variables $\left\{ \alpha(t,x^{k}),\beta_{+}(t,x^{k}),\beta_{-}(t,x^{k})\right\} $
via the transformation 
\begin{equation}
\Biggl\{\begin{array}{l}
q_{1}(t,x^{k})=\left(\alpha(t,x^{k})+\beta_{+}(t,x^{k})+\sqrt{3}\beta_{-}(t,x^{k})\right)\\
q_{2}(t,x^{k})=\left(\alpha(t,x^{k})+\beta_{+}(t,x^{k})-\sqrt{3}\beta_{-}(t,x^{k})\right)\\
q_{3}(t,x^{k})=\left(\alpha(t,x^{k})-2\beta_{+}(t,x^{k})\right)
\end{array}
\label{7.1}
\end{equation}
Imposing the canonical condition $\left(q_{a},p_{a}\right)\longrightarrow\left(\alpha,p_{\alpha};\beta_{+},p_{+};\beta_{-},p_{-}\right)$
\begin{equation}
p_{\alpha}\dot{\alpha}+p_{+}\dot{\beta}_{+}+p_{-}\dot{\beta}_{-}=\underset{a}{\sum}p_{a}\partial_{t}q_{a},
\label{7.2}
\end{equation}
we can rewrite the $f(R)$ \emph{superHamiltonian} \eqref{6.6} of the generic
cosmological problem in Misner-like variables 
\begin{equation}
\begin{split}
\mathscr{H}&_{g,JF}  = \\ &\cfrac{2e^{-\frac{3}{2}\alpha}}{\mathbf{v}}\left[\cfrac{1}{6\phi}\left(p_{+}^{2}+p_{-}^{2}\right)+\cfrac{1}{6}\pi_{\phi}^{2}\phi-\cfrac{1}{3}p_{\alpha}\pi_{\phi}+U\left(\alpha,\beta_{\pm},\phi\right)\right],
\label{7.3}
\end{split}
\end{equation}
where $\textbf{v}(x^{l})$ = $\det\left(e_{i}^{a}\right)$ and
$U\left(\alpha,\beta_{\pm},\phi\right)=\cfrac{\mathbf{v}^{2}}{4}e^{3\alpha}\left(V(\phi)-\phi\,^{3}R+2D_{i}D^{i}\phi\right)$.

It is clear that, differently from the case in General Relativity \cite{montani2011primordial} where the \emph{superHamiltonian} takes the form 
\begin{equation}
\mathcal{H}_{g}=\cfrac{2e^{-\frac{3}{2}\alpha}}{\mathbf{v}}\left(-p_{\alpha}^{2}+p_{+}^{2}+p_{-}^{2}+U(\alpha,\beta_{\pm})\right),
\label{7.4}
\end{equation}
in the case of the $f(R)$ theories in the Jordan frame we lose the
the pseudo-Riemannian structure. It is also evident that the equation
\eqref{7.3} is the classical counterpart in the cosmological context of
the Wheeler-De Witt quantum-gravitational equation \eqref{3.11}. Therefore,
the same considerations about the problem of identifying a time variable
apply. As we have already done, we highlight the linearity of the
equation with respect to the conjugated momentum $p_{\alpha}$ and,
therefore, the formal analogy between the equation \eqref{7.3} and a Schrodinger-like
equation, once one assigns to $\alpha$ the role of the timelike variable
of the theory. The presence of a potential term still does not allow
to construct a Hilbert space for physical quantum states.

\part{Classical $f(R)$ Cosmology}

When we use the term classical cosmology we refer to the description
of the Universe using the tools provided by General Relativity and,
in this case, the equivalence with Brans-Dicke theory of $f(R)$ models.
Our purpose is to analyse the Bianchi IX model, which is the prototype
of the generic cosmological solution, in Jordan frame and try to solve
some already known issues of this model in GR, such as
its chaotic behaviour as we are going through the cosmological singularity.
In order to achieve this result we need to follow some intermediate
steps: find a solution for the Bianchi I model, \emph{i.e.} the Kasner
solution, and use these solutions to discuss some properties of the
potential terms. We want to define a class of $f(R)$ theory for which
we can neglect the term $V(\phi)$ in a way that the solutions became
independent from the specific functional form of the theory. Then
we discuss the properties of the potential term of the Bianchi IX
model and try to find a region, in the space of parameters that describes
the Kasner solutions, which removes the chaos. Finally, we will discuss
the properties of our cosmological model, studying the dynamical features
of the bounces that characterize the Bianchi IX model.

\section{Kasner solution for the Bianchi I model}

The Bianchi I Universe corresponds to the case $V_{B}\left(\alpha,\beta_{\pm},\phi\right)=0$,
since the structure constants of the isometry group are all zero.
Strictly speaking, we are assuming also the potential $V\left(\phi\right)$
to be negligible. The conditions
under which the potential of the $f(R)$ theory is negligible will
be discussed later. The \emph{superHamiltonian} \eqref{7.3} becomes 
\begin{equation}
\mathcal{H}_{BI,JF}=\cfrac{2e^{-\frac{3}{2}\alpha}}{\mathbf{v}}\left[\cfrac{1}{6\phi}\left(p_{+}^{2}+p_{-}^{2}\right)+\cfrac{1}{6}\pi_{\phi}^{2}\phi-\cfrac{1}{3}p_{\alpha}\pi_{\phi}\right].
\label{8.1}
\end{equation}
In the framework of Bianchi models, the anisotropies of the Universe
$\beta_{\pm}$ represent the 2 physical degrees of freedom of Einstenian
gravity, the variable $\alpha$ represents an embedding degree of
freedom and the variable $\phi$ is a massive mode. Solving classically
the \emph{superHamiltonian} constraint \eqref{8.1} 
with respect to the conjugate momentum $\pi_{\phi}$, we get the following
quadratic equation 
\begin{equation}
\pi_{\phi}^{2}-\cfrac{2p_{\alpha}}{\phi}\pi_{\phi}+\cfrac{\left(p_{+}^{2}+p_{-}^{2}\right)}{\phi^{2}}=0,
\label{8.2}
\end{equation}
which admits the solutions 
\begin{equation}
\pi_{\phi\,_{1,2}}=\cfrac{p_{\alpha}}{\phi}\pm\cfrac{1}{\phi}\sqrt{p_{\alpha}^{2}-p_{+}^{2}-p_{-}^{2}}=-h_{\phi\,_{1,2}}.
\label{8.3}
\end{equation}

The next step in the procedure consists in the imposition of the so-called
time gauge which sets the lapse function $N$ 
\begin{equation}
\dot{\phi}=N\cfrac{\partial\mathcal{H}_{BM\,\mathrm{I},JF}}{\partial\pi_{\phi}}=\cfrac{2Ne^{-\frac{3}{2}\alpha}}{3\mathbf{v}}\left(\phi\pi_{\phi}-p_{\alpha}\right)=1,
\label{8.4}
\end{equation}
\begin{equation}
N=N_{ADM}=\cfrac{3\mathbf{v}e^{\frac{3}{2}\alpha}}{2\left(\phi\pi_{\phi}-p_{\alpha}\right)}.
\label{8.5}
\end{equation}
Since $N$ is always positive, we obtain the condition $\phi\pi_{\phi}>p_{\alpha}$
which constraints the choice of the positive sign in equation \eqref{8.3}.
From the Hamilton's equations 
\begin{equation}
\left\{ \begin{array}{l}
\cfrac{\partial\alpha}{\partial\phi}=\cfrac{\partial h_{\phi}}{\partial p_{\alpha}}=-\cfrac{1}{\phi}\left(1+\cfrac{p_{\alpha}}{\sqrt{p_{\alpha}^{2}-p_{+}^{2}-p_{-}^{2}}}\right)\\
\cfrac{\partial p_{\alpha}}{\partial\phi}=-\cfrac{\partial h_{\phi}}{\partial\alpha}=0
\end{array}\right.
\label{8.6}
\end{equation}
we can derive the equation of motion 
\begin{equation}
\alpha\left(\phi\right)=-Kln\left(\phi\right)+\alpha_{0},
\label{8.7}
\end{equation}
where we set $\alpha_{0}=0$ by choosing the initial condition $\alpha(0)=0$
and where $K$ is a constant of motion defined as 
\begin{equation}
K=1+\cfrac{p_{\alpha}}{\sqrt{p_{\alpha}^{2}-p_{+}^{2}-p_{-}^{2}}}.
\label{8.8}
\end{equation}
Since the Misner variable $\alpha$ is related to the volume of the
Universe, the constant $K$ must always be real. Consequently, we
must impose the relation 
\begin{equation}
p_{\alpha}^{2}>p_{+}^{2}+p_{-}^{2},
\label{8.9}
\end{equation}
which constraints the value of $|K|>1$. 
The Hamilton's equations takes the following form: $d\beta_{\pm}/d\phi=\partial h_{\phi}/\partial p_{\pm}$
and, since the Hamiltonian does not depend on the coordinates, the
momenta are constant. We can express the \emph{Misner coordinates}
as function of the scalar field:

\begin{equation}
\begin{split} & \alpha=-\left(1+\frac{p_{\alpha}}{\sqrt{p_{\alpha}^{2}-p_{+}^{2}-p_{-}^{2}}}\right)ln{\phi},\\
 & \beta_{\pm}=-\left(\frac{p_{\pm}}{\sqrt{p_{\alpha}^{2}-p_{+}^{2}-p_{-}^{2}}}\right)ln{\phi}.
\end{split}
\label{8.11}
\end{equation}
By dividing between $\beta_{\pm}\left(\phi\right)$ and $\alpha\left(\phi\right)$
we can find $\beta_{\pm}=\beta_{\pm}\left(\alpha\right)$

\begin{equation}
\beta_{\pm}\left(\alpha\right)=\frac{p_{\pm}}{p_{\alpha}\pm\sqrt{p_{\alpha}^{2}-p_{+}^{2}-p_{-}^{2}}}\cdot\alpha.
\label{8.12}
\end{equation}

\section{Potential term of $f(R)$ theory}

We want to deal with the most general solution, so we have to set
some constraint on the functional form of the theory. As the model
we are analysing is important only in a pre-inflation scenario , we
are interested in the limit $\alpha\rightarrow-\infty$ and so we
want that the potential term of $\phi$ in the Hamiltonian of the
system vanish as we are going to the cosmological singularity. So
we can work with a class of theories, that obeys to this constraint,
instead of dealing with a specific theory with a specific functional
form of the potential. In order to achieve our goal we will use the
\emph{kasner solutions} found in the previous section and then we
will discuss the functional form with the dominant diverging term
which admits the limit we are looking for. Finally, we will define
a concrete constraint of the functional form of the potential of the
$f(R)$ theory.

We have to rewrite the \emph{kasner solutions} because we are interested
in $\phi$ as a function of $\alpha$. The Kasner solution gives us
$\alpha(\phi)$ in eq.\eqref{8.11} that can be easily inverted to find
$\phi(\alpha)$ 
\begin{equation}
\alpha=-\left(1+\frac{p_{\alpha}}{\sqrt{p_{\alpha}^{2}-p_{+}^{2}-p_{-}^{2}}}\right)ln{\phi},
\label{9.1}
\end{equation}
and 
\begin{equation}
\phi=e^{-\frac{\alpha}{K}},
\label{9.2}
\end{equation}
whit $K=1+p_{\alpha}/\sqrt{p_{\alpha}^{2}-p_{+}^{2}-p_{-}^{2}}$.
We can assume that $V(\phi)\propto\phi^{n}$, because other dependence
are forbidden by the condition chosen on the potential.

\begin{equation}
e^{3\alpha}V(\phi)\propto e^{3\alpha}e^{-n\alpha/K}=e^{\left(3-n/K\right)\alpha},
\label{9.3}
\end{equation}
we have to study the sign of $3-n/K$ because it must be positive
in order to have a vanishing potential. We will now calculate the
maximum $n$ that respects the condition 
\begin{equation}
3-n/K=3-\frac{n}{1+\frac{p_{\alpha}}{\sqrt{p_{\alpha}^{2}-p_{+}^{2}-p_{-}^{2}}}}>0.
\label{9.4}
\end{equation}
The inequalities can be rewritten by solving it for $n$ 
\begin{equation}
n<3\left(1+\frac{p_{\alpha}}{\sqrt{p_{\alpha}^{2}-p_{+}^{2}-p_{-}^{2}}}\right).
\label{9.5}
\end{equation}
The function of the momenta is strictly increasing and its maximum
and minimum value are defined by its limits, respectively to $-\infty$
and to $+\infty$. 
\begin{equation}
-1<\frac{p_{\alpha}}{\sqrt{p_{\alpha}^{2}-p_{+}^{2}-p_{-}^{2}}}<1,
\label{9.6}
\end{equation}
and so the value of $K$ is also limited between 0 and 2. We are interested
to the maximum value of $n$ and then we get $K=2$, which gives us
the value of $n$ we are looking for 
\begin{equation}
n_{MAX}=6.
\label{9.7}
\end{equation}
For all the metric $f(R)$ theories, with a potential term which diverges
slower than $\phi^{6}$, the present and following analyses hold.

The form of the function $f$, due to a polynomial form of the potential $V(\phi)$, can be derived using the following differential equation:
\begin{equation}
    f''(R)[R-V'(f'(R))]=0.
\end{equation}
The first solution ($f(R)=R$) leads to the General Relativity, while the second solution gives us the following form of the function:
\begin{equation}
    f(R)=(n-1)\biggl(\frac{R}{n}\biggr)^{\frac{n}{n-1}}.
\end{equation}
so we can fix $n \in (1,6]$.

\section{Chaotic behaviour of the Bianchi IX potential}

We have already seen that the dynamics of the Bianchi IX model is
completely determined by the potential. The form of the potential,
in its asymptotic form, is a wall with a triangular symmetry but in
a $f(R)$ theory it is multiplied by the scalar field $\phi$. This
feature gives us the chance to remove the chaotic behaviour of this
model, characterized by oscillations from one Kasner solution to another.
We are looking for a region, in the space of the parameters that characterized
the Kasner solutions, that ensures us that the potential term will
vanish as we are going to the cosmological singularity. We analyse
the role of the scalar field using the method of consistent potential
(MCP), \emph{i.e.} we assume that the approach to the singularity
is asymptotically velocity term dominated\cite{PhysRevD.61.023508}. If this is true, the model
approaches closely to a Kasner solution. The \emph{superHamiltonian}
constraint takes the form 
\begin{equation}
\mathcal{H}_{IX,JF}=\mathcal{H}_{k}+\mathcal{H}_{v},
\label{10.1}
\end{equation}
where $\mathcal{H}_{k}$ is the kinetic term and $\mathcal{H}_{v}$
stands as 
\begin{equation}
\begin{split}2 & \mathcal{H}_{v}=\phi\biggl\{\left(e^{2\alpha+2\beta_{+}+2\sqrt{3}\beta_{-}}+e^{2\alpha+2\beta_{+}-2\sqrt{3}\beta_{-}}+e^{2\alpha-4\beta_{+}}\right)\\
 & -2\left(e^{2\alpha+2\beta_{+}}+e^{2\alpha-\beta_{+}+\sqrt{3}\beta_{-}}+e^{2\alpha-\beta_{+}-\sqrt{3}\beta_{-}}\right)\biggr\}+e^{3\alpha}V(\phi).
\end{split}
\label{10.2}
\end{equation}
In these variables, the singularity occurs as $\alpha\rightarrow-\infty$,
and as we have discussed in the previous section $e^{3\alpha}V(\phi)\rightarrow0$
going to the singularity, so we shall ignore this term. The method
of consistent potential (MCP) will be used in order to define a \emph{Kasner
stability region} in which the characteristic oscillation of the Mixmaster
Universe are suppressed after a given sequence. The MCP requires to
assume $\mathcal{H}=\mathcal{H}_{k}$. Variation of this Hamiltonian
yields equations with the solution obtained in eq.\eqref{8.11} which we are
going to express as function of $\alpha$

\begin{equation}
\begin{split} & \beta_{\pm}=\beta_{\pm}^{'}\alpha,\\
 & \phi=e^{-\frac{\alpha}{K}},
\end{split}
\label{10.3}
\end{equation}
where $K$ and $v_{\pm}$ are defined as

\begin{equation}
\begin{split} & \beta{'}_{\pm}=\frac{p_{\pm}}{\sqrt{p_{\alpha}^{2}-p_{+}^{2}-p_{-}^{2}}},\\
 & \text{\ensuremath{K}}=\left(1+\frac{p_{\alpha}}{\sqrt{p_{\alpha}^{2}-p_{+}^{2}-p_{-}^{2}}}\right).
\end{split}
\label{10.4}
\end{equation}
The minisuperspace potential is dominated by the first three terms
of r.h.s. of eq. \eqref{10.2} 
\begin{equation}
2\mathcal{H}_{v}\approx\phi\left(e^{2\alpha+2\beta_{+}+2\sqrt{3}\beta_{-}}+e^{2\alpha+2\beta_{+}-2\sqrt{3}\beta_{-}}+e^{2\alpha-4\beta_{+}}\right),
\label{10.5}
\end{equation}
so we can write $K$ as a function of $\beta'_{+}$ and $\beta'_{-}$ considering that
\[
\beta{'}_{+}^{2}+\beta{'}_{-}^{2}=\frac{p_{+}^{2}+p_{-}^{2}}{\left(p_{\alpha}+\sqrt{p_{\alpha}^{2}-p_{+}^{2}-p_{-}^{2}}\right)^{2}},
\]
and we can invert this relation to find $p_{+}^{2}+p_{-}^{2}$ 
\begin{equation}
p_{+}^{2}+p_{-}^{2}=\frac{4p_{\alpha}^{2}\left(\beta{'}_{+}^{2}+\beta{'}_{-}^{2}\right)}{\left(\beta{'}_{+}^{2}+\beta{'}_{-}^{2}-1\right)^{2}},
\label{10.6}
\end{equation}
inserting eq. \eqref{10.6} into the definition of $K$ we can express it
as function of $\beta_{+}$ and $\beta_{-}$: 
\begin{equation}
\frac{1}{2K}=\frac{\beta{'}_{+}^{2}+\beta{'}_{-}^{2}-1}{4\beta{'}_{+}^{2}+4\beta{'}_{-}^{2}}.
\label{10.7}
\end{equation}
Substitution of \eqref{10.4} and \eqref{10.7} into \eqref{10.5}, yields to 
\begin{equation}
\begin{split}2\mathcal{H}_{v}\approx & \exp\biggl\{{2\alpha\left(1+\beta{'}_{+}+\sqrt{3}\beta{'}_{-}-\frac{\beta{'}_{+}^{2}+\beta{'}_{-}^{2}-1}{4\beta{'}_{+}^{2}+4\beta{'}_{-}^{2}}\right)}\biggr\}\\
+ & \exp\biggl\{{2\alpha\left(1+\beta{'}_{+}-\sqrt{3}\beta{'}_{-}-\frac{\beta{'}_{+}^{2}+\beta{'}_{-}^{2}-1}{4\beta{'}_{+}^{2}+4\beta{'}_{-}^{2}}\right)}\biggr\}\\
+ & \exp\biggl\{{2\alpha\left(1-2\beta{'}_{+}-\frac{\beta{'}_{+}^{2}+\beta{'}_{-}^{2}-1}{4\beta{'}_{+}^{2}+4v_{-}^{2}}\right)}\biggr\}.
\end{split}
\label{10.8}
\end{equation}
We are looking for some values of $\beta_{+}$ and $\beta_{-}$ which
make all the terms going to zero as $\alpha\rightarrow-\infty$.
In order to achieve our purpose we had to solve, at the same time,
all these inequalities: 
\begin{equation}
\begin{split} & 1+\beta{'}_{+}+\sqrt{3}\beta{'}_{-}-\frac{\beta{'}_{+}^{2}+\beta{'}_{-}^{2}-1}{4\beta{'}_{+}^{2}+4\beta{'}_{-}^{2}}>0,\\
 & 1+\beta{'}_{+}-\sqrt{3}\beta{'}_{-}-\frac{\beta{'}_{+}^{2}+\beta{'}_{-}^{2}-1}{4\beta{'}_{+}^{2}+4\beta{'}_{-}^{2}}>0,\\
 & 1-2\beta{'}_{+}-\frac{\beta{'}_{+}^{2}+\beta{'}_{-}^{2}-1}{4\beta{'}_{+}^{2}+4\beta{'}_{-}^{2}}>0.
\end{split}
\label{10.9}
\end{equation}
With the help of the software \emph{Mathematica} we have solved the
three inequalities to find the values of $\beta_{+}$ and $\beta_{-}$
which can make all the three exponential to vanish at the same time.
The explicit form of the \emph{Kasner stability region} can be found
in Appendix B, now we just want to show its graph.

\begin{figure}[h]
\includegraphics[scale=0.25]{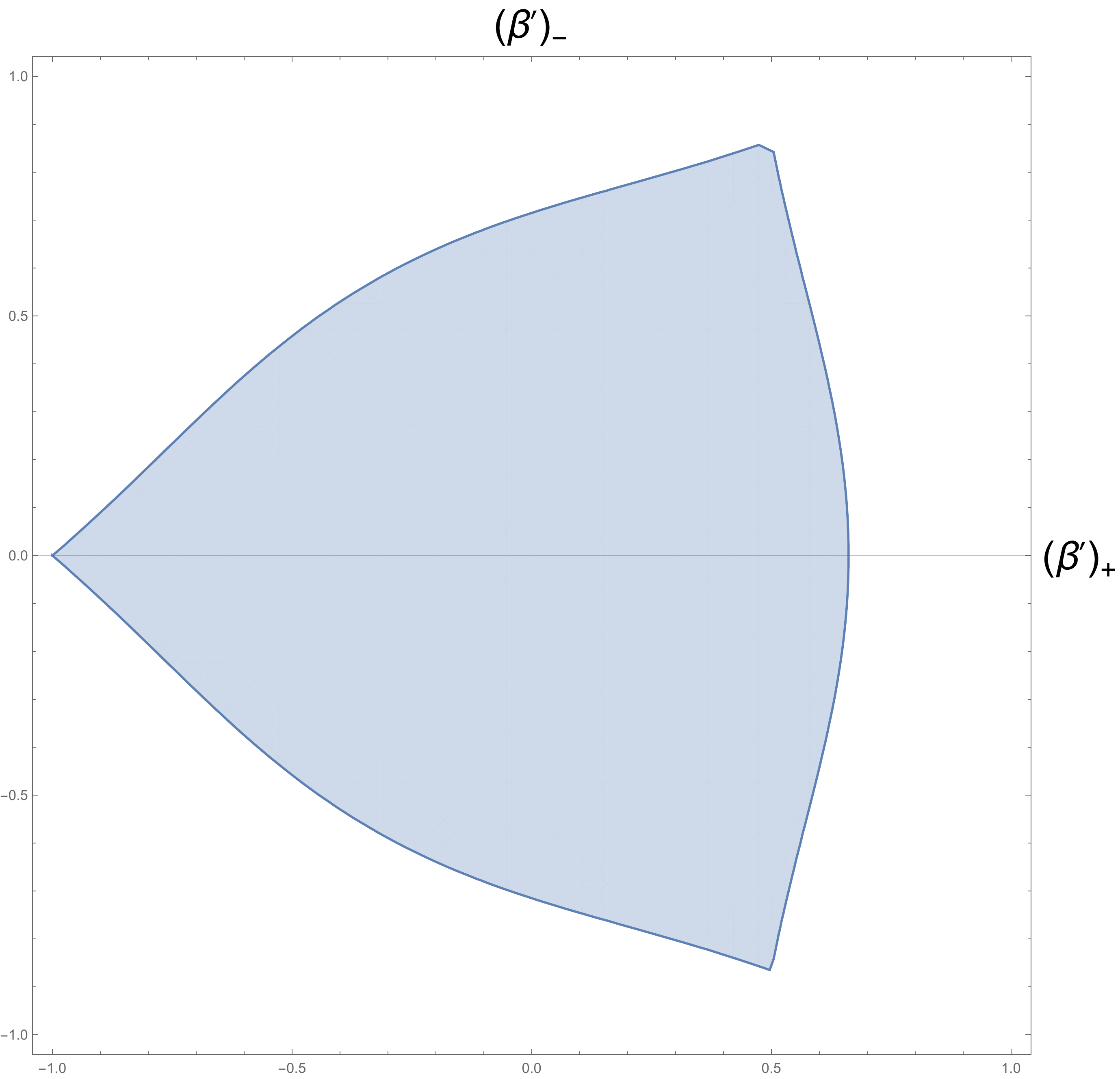}
\caption{\label{fig2}Kasner stability region as function of $\beta'_{+}$ and $\beta'_{-}$ }
\end{figure}

Now our purpose is to characterize the dynamical features of the "bounces"
typical of the the Bianchi IX model. We want to check if the
\emph{Kasner stability region} is an attractor for the Mixmaster dynamics.
First of all, we shall remember that, near the singularity, the matter
and radiation density terms are negligible because the dynamics of
the Universe is dominated by the curvature term due to the geometry
of the space-time. The potential term of the Bianchi IX cosmology
depends on the variable $\alpha$ and this feature complicates the
dynamics with respect to the \emph{Kasner solution}, generating in
principle a chaotic evolution. The vacuum Bianchi IX dynamics is constituted
by an infinite series of Kasner regimes, associated with the continuous
scatterings against the infinite potential walls. The purpose of this
chapter is to verify if, thanks to the additional degree of freedom
introduced by the choice of $f(R)$ framework, we are led to remove
the chaotic behaviour. We have already find a region, in the space
 of the parameters $(\beta_{+},\beta_{-})$ that describe the Kasner
solutions, for which the point-Universe does not impact against the
potential walls. Now we want to verify if the natural evolution of
the system, starting from any point in the parameters space, will
bring the point-Universe to the \emph{Kasner stability region}. The
calculus of the equation of motion is very complicated because of
the potential term, so we will not integrate numerically the equation
of motion (which are not clearly analytical integrable) to determinate
the evolution of the cosmological model. Instead, recalling the method
used to find the eq.\eqref{5.9} we will search for three constants of motion
that allow us to find the new Kasner solution, after the bounce, as
a function of the previous solution. The \emph{superHamiltonian }constraint
takes the form: 
\begin{equation}
\mathcal{H}_{IX,JF}=\phi\pi_{\phi}^{2}-2\pi_{\phi}p_{\alpha}+\frac{p_{+}^{2}+p_{-}^{2}}{\phi}+\phi e^{3\alpha}V_{IX}(\alpha,\beta_{\pm})=0.
\label{10.10}
\end{equation}
Where the potential of the model is 
\begin{equation}
\begin{split}V_{IX}(\alpha, & \beta_{\pm})=\\
 & \frac{1}{2}e^{-3\alpha}\biggl\{\left(e^{2\alpha+2\beta_{+}+2\sqrt{3}\beta_{-}}+e^{2\alpha+2\beta_{+}-2\sqrt{3}\beta_{-}}+e^{2\alpha-4\beta_{+}}\right)\\
 & -\left(e^{2\alpha+2\beta_{+}}+e^{2\alpha-\beta_{+}+\sqrt{3}\beta_{-}}+e^{2\alpha-\beta_{+}-\sqrt{3}\beta_{-}}\right)\biggr\},
\end{split}
\label{10.11}
\end{equation}
where the first three terms dominate as we are going to the singularity.
The potential $V(\phi)$ can be neglected, as we saw previously. We
will solve the \emph{superHamiltonian} constraint with respect to
the appropriate time-like variable $\phi$:

\begin{equation}
\pi_{\phi}=-h_{\phi}=\frac{1}{\phi}\left\{ p_{\alpha}+\sqrt{p_{\alpha}^{2}+p_{+}^{2}+p_{-}^{2}-6\phi e^{3\alpha}V_{IX}(\alpha,\beta_{\pm})}\right\}.
\label{10.12}
\end{equation}
In the standard case, the evolution of the Universe is described by
giving $\beta_{\pm}$ as functions of the scalar field $\phi$. The
entire problem is governed by the function $V_{IX}(\alpha,\beta_{\pm})$,
which has the symmetry of an equilateral triangle as in FIG.\ref{fig1}.
For $\beta_{+}\rightarrow-\infty$ it gets the following asymptotic
form 
\begin{equation}
V_{IX}\sim\frac{1}{3}e^{2\alpha-4\beta_{+}}
\label{10.13}
\end{equation}
showing one of the three exponentially steep walls on which the equipotentials
are straight lines. The corners of this triangular potential are open;
it satisfies the condition $V_{IX}\ge0$ and vanishes only at the origin,
where $V_{IX}\approx2\left(\beta_{+}^{2}+\beta_{-}^{2}\right)$. Because
the potential rises so steeply for large $\beta$, little time is
spent with the $\beta$ bouncing against the potential wall and most
of the time is spent in free motion when $V_{IX}$ can be neglected, so
the motion is described by a Kasner solution: 
\begin{equation}
\begin{split} & \alpha=-\left(1+\frac{p_{\alpha}}{\sqrt{p_{\alpha}^{2}-p_{+}^{2}-p_{-}^{2}}}\right)ln{\phi},\\
 & \beta_{\pm}=\left(\frac{p_{\pm}}{\sqrt{p_{\alpha}^{2}-p_{+}^{2}-p_{-}^{2}}}\right)ln{\phi}.
\end{split}
\label{10.14}
\end{equation}
By defining the velocities as the derivatives with respect to the
logarithm of scalar field 
\begin{equation}
\begin{split} & \alpha'=1+\frac{p_{\alpha}}{\sqrt{p_{\alpha}^{2}-p_{+}^{2}-p_{-}^{2}}},\\
 & v_{\pm}=\frac{p_{\pm}}{\sqrt{p_{\alpha}^{2}-p_{+}^{2}-p_{-}^{2}}},
\end{split}
\label{10.15}
\end{equation}
with the condition $(\alpha'-1)^{2}-(\beta_{+}')^{2}-(\beta_{-}')^{2}=1$.
For the sake of simplicity we study the bouncing against the potential
of an assigned Kasner solution, considering the vertical wall which
intersects the axes $\beta_{+}$. We recall that the choice of this
wall is equivalent to anyother one, due to the traingular simmetry
of the potential (under a rotation of $\frac{\pi}{3}$). We use the
asymptotic form of the \emph{superHamiltonian} using eq.\eqref{10.13} 
\begin{equation}
\mathcal{H}_{IX,JF}=\frac{1}{\phi}\left\{ p_{\alpha}+\sqrt{p_{\alpha}^{2}-p_{+}^{2}-p_{-}^{2}+\frac{1}{3}e^{+2\alpha-4\beta_{+}}}\right\}.
\label{10.16} 
\end{equation}
It is independent of $\beta_{-}$ in this approximation, so $p_{-}$
will be constant during the bounce. Another constant of motion can
be found by comparing these equations 
\begin{equation}
\begin{split} & p_{+}'=-\frac{\partial\mathcal{H}_{IX,JF}}{\partial\beta_{+}}=4\mathcal{H}_{IX,JF}^{-1}e^{2\alpha-4\beta_{+}},\\
 & p_{\alpha}'=-\frac{\partial\mathcal{H}_{IX,JF}}{\partial\alpha}=-2\mathcal{H}_{IX,JF}^{-1}e^{2\alpha-4\beta_{+}},
\end{split}
\label{10.17}
\end{equation}
with the results that $\frac{1}{2}p_{+}+p_{\alpha}=const$. A third
constant of motion can be found in the same way, by comparing the
two terms $\frac{\partial\phi\pi_{\phi}}{\partial\phi}$ and $\frac{\partial p_{\alpha}}{\partial\phi}$
, so we have $\phi\pi_{\phi}-1/2p_{\alpha}=const$. Thus, the system
of equations that we want to solve is: 
\begin{equation}
\left\{ \begin{array}{l}
1/2p_{\alpha}^{'}+\sqrt{p_{\alpha}^{2'}-p_{+}^{2'}-p_{-}^{2'}}=1/2p_{\alpha}+\sqrt{p_{\alpha}^{2}-p_{+}^{2}-p_{-}^{2}},\\
\frac{1}{2}p_{+}^{'}+p_{\alpha}^{'}=\frac{1}{2}p_{+}+p_{\alpha},\\
p_{-}^{'}=p_{-},
\end{array}\right.
\label{10.19}
\end{equation}
where the primate quantities are the ones after the bounce. In order
to solve the system we use the Kasner solution to write the equations
as functions of $v_{\pm},v_{\alpha}$. The velocities satisfy the condition: $v_{\alpha}^2-v_+^2-v_-^2=1$, which suggest the following change of variables:
\begin{equation}
\begin{split}
&v_{\alpha}=\cosh{\varphi},\\
&v_+=\sinh{\varphi}\cos{\theta},\\
&v_-=\sinh{\varphi}\sin{\theta}. 
\end{split}
\label{10.20}
\end{equation} 

Far from the potential walls,
the Hamiltonian is independent from the coordinates and so the conjugate
momenta are constants. We can rewrite the constant of motion as function of $\varphi$ and $\theta$:

\begin{equation}
\left\{ \begin{array}{l}
\mathcal{H}_{IX,JF}\left(\frac{1}{2}\cosh{\varphi}+1\right),\\
\mathcal{H}_{IX,JF}\left(\frac{1}{2}\sinh{\varphi}\cos{\theta}+\cosh{\varphi}\right),\\
\mathcal{H}_{IX,JF}\sinh{\varphi}\sin{\theta}.
\end{array}\right\}
\label{10.21}
\end{equation}

We have developed a software to help us to simulate the evolution
of the system. We have performed many simulations to understand if
the system was able to reach the stability region starting from any
point in the space of the parameters that describes the Kasner solutions
$(\beta_{+}^{'},\beta_{-}^{'})$. Since $|\beta_{\pm}^{'}|<1$ we
choose to evolve 10.000 models inside 4 circles on the four corners
of the square with a radius $1/10$.

\begin{figure}[h]
\centering
\subfloat[][\emph{Initial positions}.]
{\includegraphics[width=.35\columnwidth]{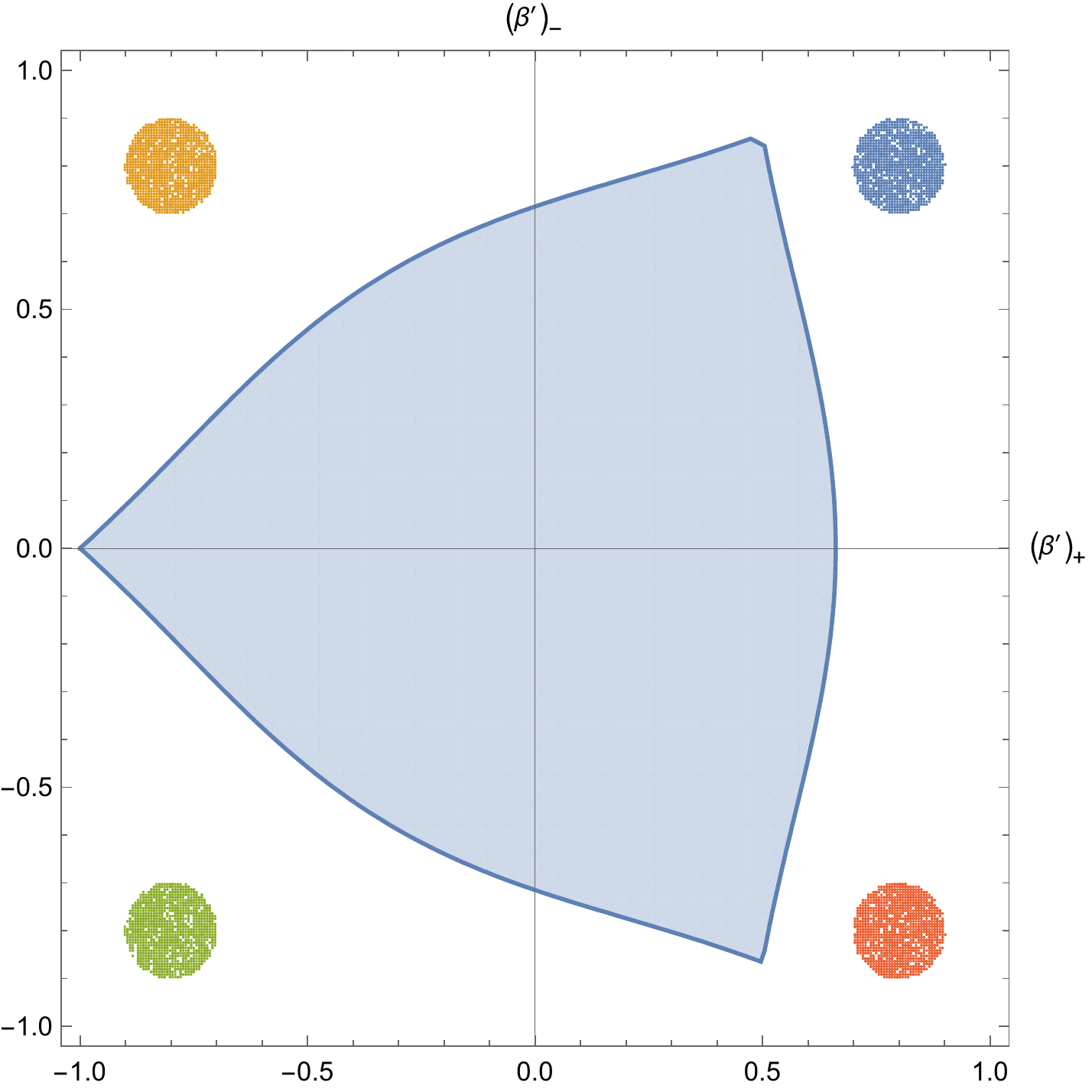}} \quad
\subfloat[][\emph{After the first bounce}.]
{\includegraphics[width=.35\columnwidth]{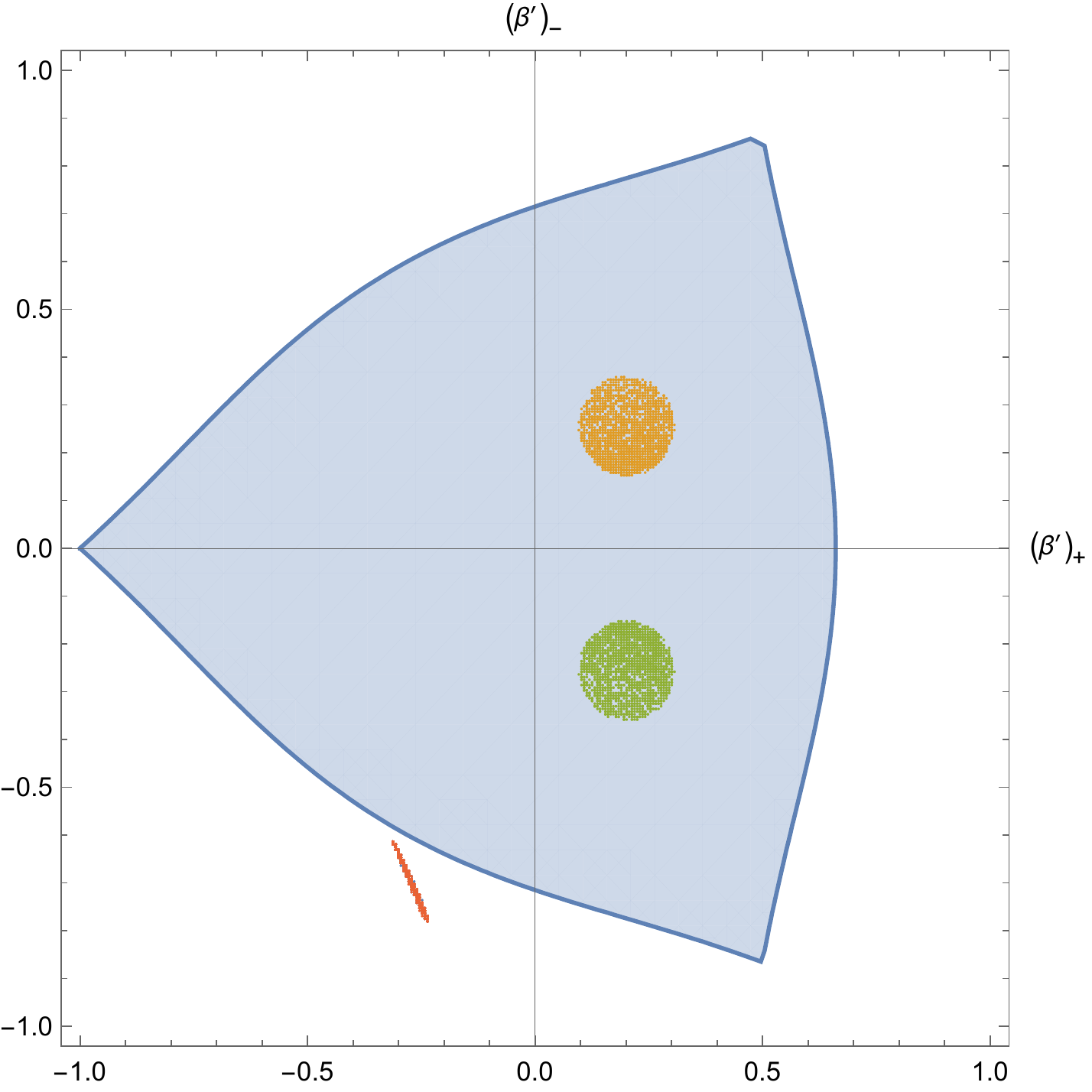}} \\
\subfloat[][\emph{After the second bounce}.]
{\includegraphics[width=.35\columnwidth]{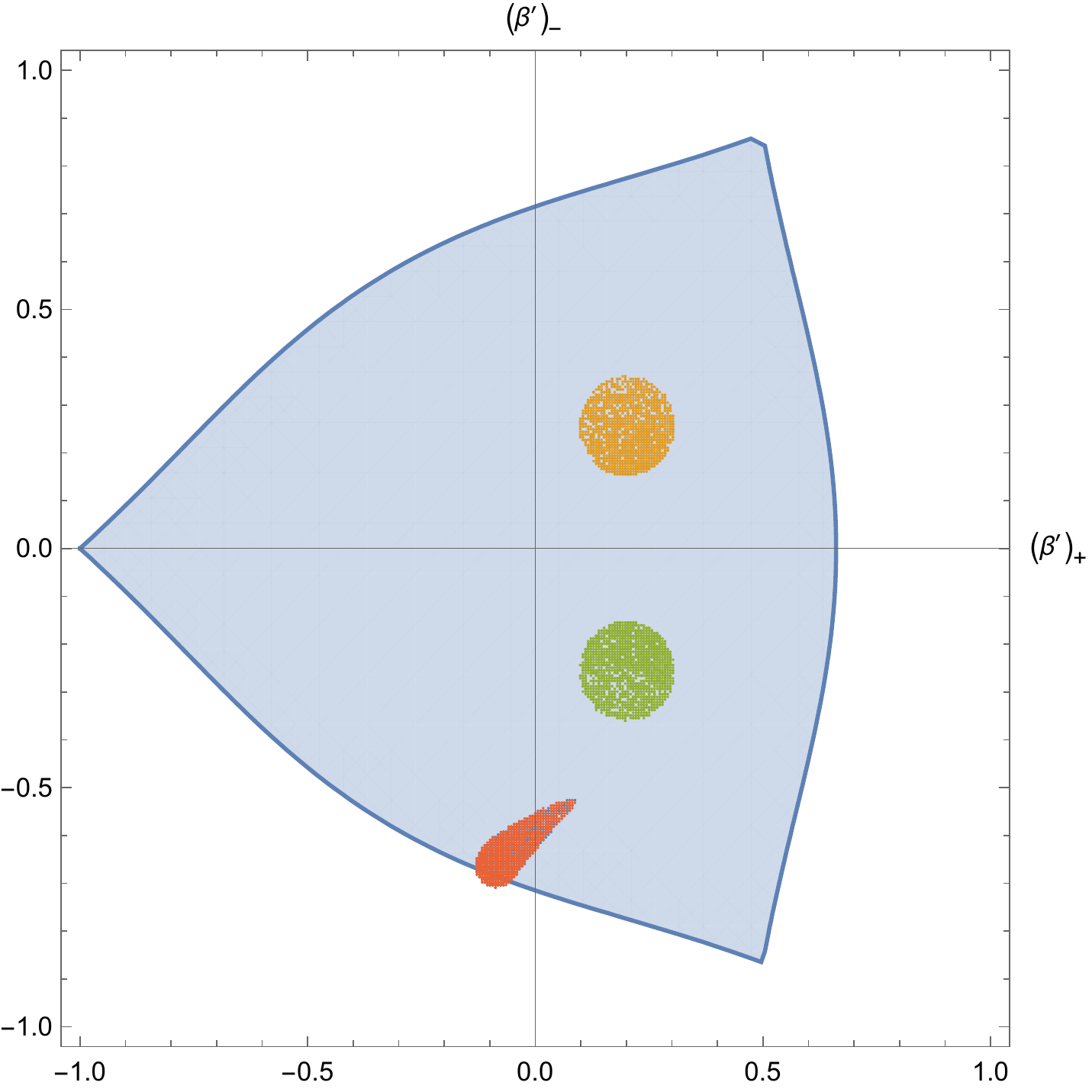}} \quad
\subfloat[][\emph{After the third bounce}.]
{\includegraphics[width=.35\columnwidth]{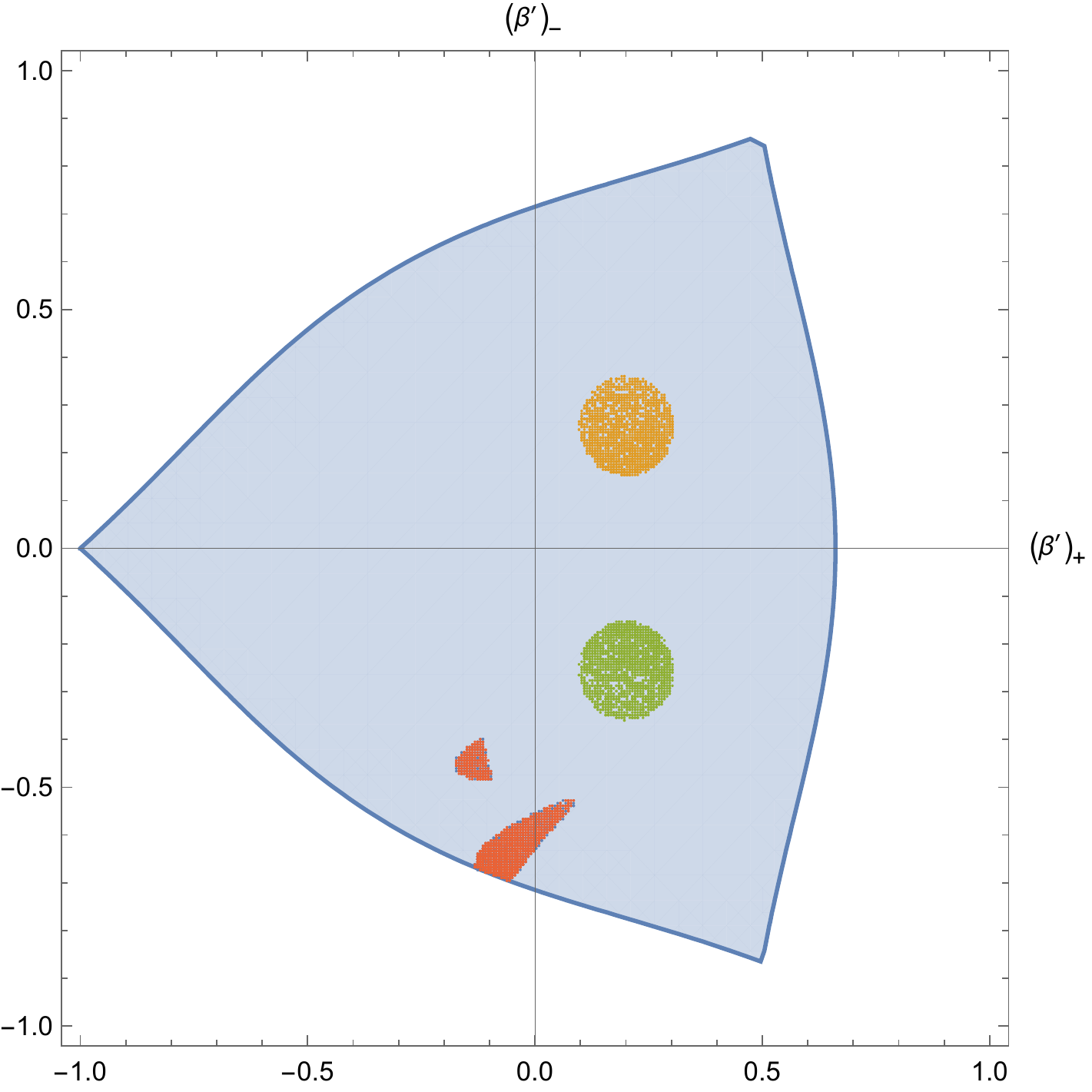}}
\caption{\label{fig3}The evolution of $10^4$ models through the map, after three bounces all the solutions has reached the stability region.}
\end{figure}

The constants of motion are written as functions of $v_{+}$ and $v_{-}$
while the \emph{Kasner stability region} is a function of $\beta_{\pm}^{'}$,
we can write $\beta_{\pm}^{'}=\frac{v_{\pm}}{v_{\alpha}+1}$ and evaluate
the position of the model in the Kasner parameter' space after every
bounce against the potential walls. Every run of the model, started
with random parameters inside the circles, has reached the stability
region and so we can deduce that the natural evolution of a Bianchi
IX cosmology in the Jordan frame naturally removes chaos.

\part{Quantum $f(R)$ Cosmology }

With the term quantum cosmology (QC) we refer to the application
of the quantum theory of gravity to the entire Universe. The existence
of such a theory would clarify the physics of the Big Bang, describing
the entire Universe like a relativistic-quantum object.

\section{Bianchi I Universe}

By promoting the \emph{superHamiltonian} constraint \eqref{8.1} to a
quantum operator annihilating the wave function $\varphi\left(\alpha,\beta_{\pm},\phi\right)$,
we obtain the Wheeler-De Witt equation 
\begin{equation}
\begin{array}{l}
\mathcal{\hat{H}}_{B\,\mathrm{I},JF}|\varphi\rangle\\
=\cfrac{2e^{-\frac{3}{2}\alpha}}{\mathbf{v}}\left[\cfrac{1}{6\phi}\left(\hat{p}_{+}^{2}+\hat{p}_{-}^{2}\right)+\cfrac{1}{6}\hat{\pi}_{\phi}^{2}\phi-\cfrac{1}{3}\hat{p}_{\alpha}\hat{\pi}_{\phi}\right]|\varphi\rangle\\
=\cfrac{2e^{-\frac{3}{2}\alpha}}{\mathbf{v}}\hbar^{2}\left[\cfrac{1}{3}\partial_{\alpha}\partial_{\phi}-\cfrac{1}{6\phi}\left(\partial_{+}^{2}+\partial_{-}^{2}\right)-\cfrac{\phi}{6}\partial_{\phi}^{2}\right]\varphi=0.
\end{array}
\label{11.1}
\end{equation}
In the kinetic term mixed impulses in the variables $\left(\alpha,\phi\right)$
appear and consequently, it is impossible to trace the formalism back
to the problem of a free relativistic particle. In the following,
we will assume $\phi$ as a timelike variable. Such a choice, in the
Jordan frame, is interesting for the idea of using a gravitational
degree of freedom as a quantum time. In this way, the problem of time
in quantum gravity could be addressed through a generalization of
General Relativity. We choose a particular factor ordering of the
Wheeler-De Witt equation rewriting the equation as 
\begin{equation}
\cfrac{2e^{-\frac{3}{2}\alpha}}{\mathbf{v}}\hbar^{2}\left[\cfrac{1}{3}\partial_{\alpha}\partial_{\phi}-\cfrac{1}{6\phi}\left(\partial_{+}^{2}+\partial_{-}^{2}\right)-\cfrac{1}{6}\partial_{\phi}\left(\phi\partial_{\phi}\right)\right]\varphi=0.
\label{11.2}
\end{equation}
Substituting in eq.\eqref{11.2} the natural solution 
\begin{equation}
\varphi\left(\alpha,\beta_{\pm},\phi\right)=Ae^{i\left(k_{+}\beta_{+}+k_{-}\beta_{-}+k_{\alpha}\alpha\right)}f(\phi),
\label{11.3}
\end{equation}
we obtain a Bessel differential equation for the function $f\left(\phi\right)$
\begin{equation}
\phi^{2}\cfrac{\partial^{2}f}{\partial\phi^{2}}+\phi\cfrac{\partial f}{\partial\phi}\left(1-2ik_{\alpha}\right)-\left(k_{+}^{2}+k_{-}^{2}\right)f=0,
\label{11.4}
\end{equation}
with solution 
\begin{equation}
f(\phi)=c_{1}\phi^{i\left(k_{\alpha}-\sqrt{k_{\alpha}^{2}-k_{+}^{2}-k_{-}^{2}}\right)}\left(c_{2}+\phi^{2i\sqrt{k_{\alpha}^{2}-k_{+}^{2}-k_{-}^{2}}}\right)
\label{11.5},
\end{equation}
where $(c_{1},c_{2})$ are arbitrary constants to be fixed by the
initial conditions. One can easily note that, by choosing the variable
$\phi$ as the internal time of the theory, the frequency separation
procedure is unfeasible due to the lack of the term $e^{ik_{\phi}\phi}$.
Since the WDW equation \eqref{11.2} is linear, the superposition principle
holds , so the general solution, describing the quantum dynamics of
the Bianchi I Universe, admits the following Fourier representation
\begin{equation}
\Phi\left(\alpha,\beta_{\pm},\phi\right)=\int_{-\infty}^{+\infty}\int_{-\infty}^{+\infty}dk_{\alpha}dk_{+}dk_{-}A\left(k_{\alpha},k_{\pm}\right)\varphi\left(\alpha,\beta_{\pm},\phi\right),
\label{11.6}
\end{equation}
where we choose 
\begin{equation}
\begin{array}{l}
A\left(k_{\alpha},k_{\pm}\right)\\
=\cfrac{1}{\sqrt{\left(2\pi\right)^{3}}\sigma_{\alpha}\sigma_{+}\sigma_{-}}\;e^{-\left(\frac{\left(k_{\alpha}-\bar{k}_{\alpha}\right)^{2}}{2\sigma_{\alpha}^{2}}+\frac{\left(k_{+}-\bar{k}_{+}\right)^{2}}{2\sigma_{+}^{2}}+\frac{\left(k_{-}-\bar{k}_{-}\right)^{2}}{2\sigma_{-}^{2}}\right)},
\end{array}
\label{11.7}
\end{equation}
as a Gaussian probability distribution, assuming we start at the initial
time $\phi_{0}$ with a Gaussian wave packet.

\subsection{The probability density}

\begin{figure}[h]
\begin{centering}
\includegraphics[scale=0.30]{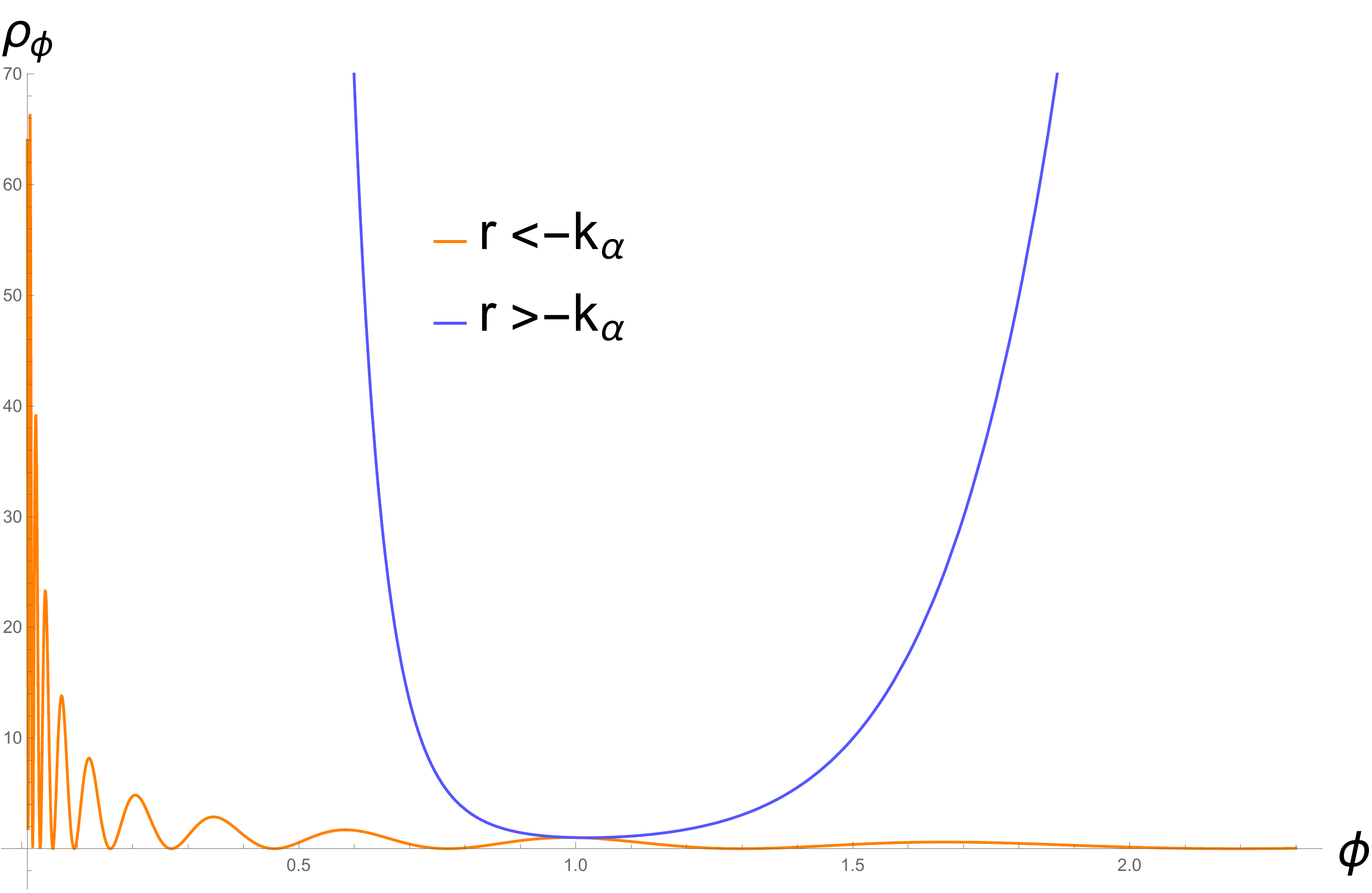} 
\par\end{centering}
\caption{\label{fig4}The trend of the probability density $\rho_{\phi}$
in function of the time $\phi$. The oscillatory $cos^{2}\left[ln\left(\phi\right)\right]$
regime is shown in orange; the hyperbolic $cosh^{2}\left[ln\left(\phi\right)\right]$
regime is shown in blue.}
\end{figure}

In order to define a Hilbert space, we must introduce a positive-defined
scalar product and, therefore, a positive-defined probability that
is preserved over time. We can build the scalar product induced by
the Wheeler-De Witt equation \eqref{11.2} by imposing the relation 
\begin{equation}
\varphi^{*}\eqref{11.2}-\varphi\eqref{11.2}^{*}=0,
\label{11.8}
\end{equation}
where $\varphi^{*}$ is the complex conjugate wave function and $\eqref{11.2}^{*}$
represents the complex conjugate Wheeler-De Witt equation. In particular,
the relation \eqref{11.8} explicitates as 
\begin{equation}
\begin{array}{l}
\cfrac{1}{6\phi}\left(\varphi^{*}\partial_{+}^{2}\varphi-\varphi\partial_{+}^{2}\varphi^{*}+\varphi^{*}\partial_{-}^{2}\varphi-\varphi\partial_{-}^{2}\varphi^{*}\right)+\\
+\cfrac{1}{6}\left(\varphi^{*}\partial_{\phi}\left(\phi\partial_{\phi}\varphi\right)-\varphi\partial_{\phi}\left(\phi\partial_{\phi}\varphi^{*}\right)\right)+\\
-\cfrac{1}{3}\left(\varphi^{*}\partial_{\alpha}\partial_{\phi}\varphi-\varphi\partial_{\alpha}\partial_{\phi}\varphi^{*}\right)=0.
\end{array}
\label{11.9}
\end{equation}
In order to search for a probability density, we want to reconduce
the quantity \eqref{11.9} to a continuity equation 
\begin{equation}
\begin{array}{ccc}
\partial_{\mu}J^{\mu}=0, &  & \mu=\left(\alpha,\beta_{+},\beta_{-},\phi\right)\end{array},
\label{11.10}
\end{equation}
We emphasize that we have relaxed the request to use a covariant four-divergence,
in favor of a Minkowskian one. Considering the so-called \emph{minisupermetric}
$g^{\mu\nu}$ in the equation \eqref{11.2}
\begin{equation}
g^{\mu\nu}=\left(\begin{array}{cccc}
0 & 0 & 0 & -\frac{1}{6}\\
0 & \frac{1}{6\phi} & 0 & 0\\
0 & 0 & \frac{1}{6\phi} & 0\\
-\frac{1}{6} & 0 & 0 & \frac{\phi}{6}
\end{array}\right)
\label{11.11}
\end{equation}
we can explicitate the relation \eqref{11.10} as 
\begin{equation}
\begin{array}{l}
\partial_{\mu}J^{\mu}=\\
\cfrac{1}{6\phi}\left(\partial_{+}J_{+}+\partial_{-}J_{-}\right)+\cfrac{1}{6}J_{\phi}\\
+\cfrac{\phi}{6}\partial_{\phi}J_{\phi}-\cfrac{1}{6}\partial_{\alpha}J_{\phi}-\cfrac{1}{6}\partial_{\phi}J_{\alpha}.
\end{array}
\label{11.12}
\end{equation}
By comparing equations \eqref{11.9} and \eqref{11.12}, the components of the conserved
four-current $J^{\mu}$ take the form 
\begin{equation}
\left\{ \begin{array}{c}
J_{\pm}=\left(\varphi^{*}\partial_{\pm}\varphi-\varphi\partial_{\pm}\varphi^{*}\right)\\
J_{\phi}=\left(\varphi^{*}\partial_{\phi}\varphi-\varphi\partial_{\phi}\varphi^{*}\right)\\
J_{\alpha}=\left(\varphi^{*}\partial_{\alpha}\varphi-\varphi\partial_{\alpha}\varphi^{*}\right)
\end{array}\right.
\label{11.13}
\end{equation}
and, in analogy with the Klein-Gordon theory, the quantity 
\begin{equation}
\int_{V_{\infty}}iJ_{\phi}d\beta_{\pm}d\alpha=\langle\varphi|\varphi\rangle,
\label{11.14}
\end{equation}
turns out to be a good candidate for a probabilty. However, this scalar
product is not positive-defined and, unlike the case of the Klein-Gordon
theory, the separation of frequencies cannot be performed. We now
define the probability density 
\begin{equation}
\begin{array}{cl}
\rho_{\phi} & =i\left(\varphi^{*}\partial_{\phi}\varphi-\varphi\partial_{\phi}\varphi^{*}\right)\\
 & =i\left(f^{*}\partial_{\phi}f-f\partial_{\phi}f^{*}\right).
\end{array}
\label{11.15}
\end{equation}
Considering the function $f$ in eq.\eqref{11.5}
\begin{equation}
\begin{array}{cc}
f\left(\phi\right) & =\phi^{i\left(k_{\alpha}-\sqrt{k_{\alpha}^{2}-k_{+}^{2}-k_{-}^{2}}\right)}\left(1+\phi^{2i\sqrt{k_{\alpha}^{2}-k_{+}^{2}-k_{-}^{2}}}\right)\\
 & =\phi^{ik_{\alpha}}\left(\phi^{-i\sqrt{k_{\alpha}^{2}-k_{+}^{2}-k_{-}^{2}}}+\phi^{i\sqrt{k_{\alpha}^{2}-k_{+}^{2}-k_{-}^{2}}}\right),
\end{array}
\label{11.16}
\end{equation}
with the choice $c_{1}=c_{2}=1$, the probability density takes the
form 
\begin{equation}
\rho_{\phi}=-\cfrac{2k_{\alpha}}{\phi}\left(2+\phi^{-2iR}+\phi^{2iR}\right),
\label{11.17}
\end{equation}
where $R=\sqrt{k_{\alpha}^{2}-k_{+}^{2}-k_{-}^{2}}$. Doing some math,
we arrange the term in parentheses 
\begin{equation}
\begin{array}{l}
\left(2+\phi^{-2iR}+\phi^{2iR}\right)=\\
=2+e^{ln\left(\phi^{2iR}\right)}+e^{ln\left(\phi^{-2iR}\right)}=\\
=2+\left(e^{i2Rln\left(\phi\right)}+e^{-i2Rln\left(\phi\right)}\right)\\
=2+2cos\left(2Rln\left(\phi\right)\right)\\
=2\left[1+cos\left(2Rln\left(\phi\right)\right)\right]=2\left[2cos^{2}\left(\cfrac{2Rln\left(\phi\right)}{2}\right)\right].
\end{array}
\label{11.18}
\end{equation}
Finally, we can rewrite the probability density \eqref{11.17} as 
\begin{equation}
\rho_{\phi}=-\cfrac{8k_{\alpha}}{\phi}\,cos^{2}\left(ln\left(\phi\right)\sqrt{k_{\alpha}^{2}-k_{+}^{2}-k_{-}^{2}}\right).
\label{11.19}
\end{equation}
It is easy to see that $\rho_{\phi}\in\mathbb{R}$ and imposing the
condition $\rho_{\phi}>0$, we find the constraint $k_{\alpha}<0$.
Using polar coordinates 
\begin{equation}
\left\{ \begin{array}{c}
k_{+}=rsin\theta\\
k_{-}=rcos\theta
\end{array}\right.
\label{11.20}
\end{equation}
we can find two trends for the probability density function, depending
on whether the term $\sqrt{k_{\alpha}^{2}-k_{+}^{2}-k_{-}^{2}}=\sqrt{k_{\alpha}^{2}-r^{2}}$
is real or imaginary (FIG.\ref{fig4}). 

In particular, being $r>0$ and $k_{\alpha}<0$, 
\begin{equation}
\begin{cases}
\textnormal{if }r<-k_{\alpha} & \textnormal{oscillatory co\ensuremath{s^{2}\left[ln\left(\phi\right)\right]} regime}\\
\textnormal{if }r>-k_{\alpha} & \textnormal{hyperbolic cos\ensuremath{h^{2}}}\left[ln\left(\phi\right)\right]\text{regime}
\end{cases}
\label{11.21}
\end{equation}
Considering the condition \eqref{8.9} derived from the Hamilton's equations
and the relation $p=\hbar k$, we conclude that only the oscillatory
regime is physically acceptable. Therefore, the wave packet \eqref{11.6}
can be rewritten as 
\begin{equation}
\begin{array}{l}
\Phi\left(\alpha,\beta_{\pm},\phi\right)\\
=\int_{-\infty}^{\;0}\int_{0}^{-k_{\alpha}}\int_{0}^{2\pi}dk_{\alpha}drd\theta\;\left[rA\left(k_{\alpha},r,\theta\right)\varphi\left(\alpha,\beta_{\pm},\phi\right)\right],
\end{array}
\label{11.22}
\end{equation}
where 
\begin{equation}
\begin{array}{l}
A\left(k_{\alpha},r,\theta\right)\\
=\cfrac{1}{\sqrt{\left(2\pi\right)^{3}}\sigma_{\alpha}\sigma_{+}\sigma_{-}}\;e^{-\left(\frac{\left(k_{\alpha}-\overline{k}_{\alpha}\right)^{2}}{2\sigma_{\alpha}^{2}}+\frac{\left(rsin\theta-\overline{rsin\theta}\right)^{2}}{2\sigma_{+}^{2}}+\frac{\left(rcos\theta-\overline{rcos\theta}\right)^{2}}{2\sigma_{-}^{2}}\right)}.
\end{array}
\label{11.23}
\end{equation}
Consequently, the normalizable probability density is 
\begin{equation}
\rho_{\phi}=i\left(\Phi^{*}\partial_{\phi}\Phi-\Phi\partial_{\phi}\Phi^{*}\right).
\label{11.24}
\end{equation}

\section{The isotropic and homogeneous FLRW Universe}

In this section, we will analyze the quantum behavior of the FLRW
Universe in the context of the $f(R)$ gravity in the Jordan frame.
In the assumption $\beta_{\pm}=0$ (and hence $p_{\pm}=0$) the \emph{superHamiltonian}
\eqref{8.1} reduces to the simpler form

\begin{equation}
\mathcal{H}_{FLRW,\,JF}=\cfrac{2e^{-\frac{3}{2}\alpha}}{\mathbf{v}}\left[\cfrac{1}{6}\pi_{\phi}^{2}\phi-\cfrac{1}{3}p_{\alpha}\pi_{\phi}\right],
\label{12.1}
\end{equation}
and quantum dynamics is described by the Wheeler-De Witt equation
\begin{align}
\cfrac{2e^{-\frac{3}{2}\alpha}}{\mathbf{v}}\hbar^{2}\left[-\cfrac{1}{6}\partial_{\phi}\left(\phi\partial_{\phi}\right)+\cfrac{1}{3}\partial_{\alpha}\partial_{\phi}\right]\varphi\left(\alpha,\phi\right)=0.
\label{12.2}
\end{align}
We note that the quantization of a homogeneous and isotropic model
has no clear physical meaning, since we lost the two Einstenian gravitational
degrees of freedom. In this respect, we can infer that the quantization
of the FLRW Universe must be mainly regarded as a toy model on which
the different quantum cosmology approaches can be easily tested. Considering
the natural solution 
\begin{equation}
\varphi=Ae^{ik_{\alpha}\alpha}f(\phi),
\label{12.3}
\end{equation}
we obtain a second-order differential equation for the function $f(\phi)$
\begin{equation}
\phi\cfrac{\partial^{2}f}{\partial\phi^{2}}+\cfrac{\partial f}{\partial\phi}\left(1-2ik_{\alpha}\right)=0,
\label{12.4}
\end{equation}
having as solution 
\begin{equation}
f(\phi)=c_{1}+c_{2}\left(\cfrac{1}{2ik_{\alpha}}\right)\phi^{2ik_{\alpha}}.
\label{12.5}
\end{equation}

\subsection{Time before quantization: the ADM reduction method}

In the model under examination, we consider the variable $\phi$ as
the embedding variable and the Misner variable $\alpha$ as the physical
degree of freedom. Solving classically the \emph{superHamiltonian}
constraint with respect to the conjugate momentum $\pi_{\phi}$ we
derive the equation 
\begin{equation}
\pi_{\phi}\left(\cfrac{1}{6}\pi_{\phi}\phi-\cfrac{1}{3}p_{\alpha}\right)=0,
\label{12.6}
\end{equation}
having as solutions 
\begin{equation}
\left\{ \begin{array}{l}
\pi_{\phi,1}=0\\
\pi_{\phi,2}=\cfrac{2p_{\alpha}}{\phi}
\end{array}\right.
\label{12.7}
\end{equation}
Considering the non-trivial one, we can define 
\begin{equation}
h_{\phi}=-\cfrac{2p_{\alpha}}{\phi},
\label{12.8}
\end{equation}
as the physical Hamiltonian that regulates the dynamics with respect
to the time $\phi$. The next step consists in the imposition of the
so-called time gauge 
\begin{equation}
\begin{array}{cl}
\dot{\phi} & =N\cfrac{\partial\mathcal{H}_{FLRW,\,JF}}{\partial\pi_{\phi}}\\
 & =\cfrac{2Ne^{-\frac{3}{2}\alpha}}{\mathbf{v}}\left(\cfrac{1}{3}\pi_{\phi}\phi-\cfrac{1}{3}p_{\alpha}\right)=1,
\end{array}
\label{12.9}
\end{equation}
which sets the the lapse function $N$ 
\begin{equation}
N=N_{ADM}=\cfrac{3\mathbf{v}e^{\frac{3}{2}\alpha}}{2\left(\pi_{\phi}\phi-p_{\alpha}\right)},
\label{12.10}
\end{equation}
and allows us to write the reduced action 
\begin{equation}
S_{red}=\int d\phi\left(p_{\alpha}\cfrac{\partial\alpha}{\partial\phi}-h_{\phi}\right),
\label{12.11}
\end{equation}
The Hamilton's equations are defined as 
\begin{equation}
\left\{ \begin{array}{c}
\cfrac{\partial\alpha}{\partial\phi}=\cfrac{\partial h_{\phi}}{\partial p_{\alpha}}=-\cfrac{2}{\phi}\\
\cfrac{\partial p_{\alpha}}{\partial\phi}=-\cfrac{\partial h_{\phi}}{\partial\alpha}=0
\end{array}\right.
\label{12.12}
\end{equation}
from which we obtain the classical trajectory with respect to time
$\phi$ 
\begin{equation}
\alpha(\phi)=-2ln\left(\phi\right)+\alpha_{0}=-2ln\left(\phi\right),
\label{12.13}
\end{equation}
where we have set $\alpha_{0}=\alpha(0)=0$.

\subsection{The probability density and the comparison between the classical
trajectory and the quantum evolution}

In analogy to the case of the Bianchi I Universe we want to define
a Hilbert space through the introduction of the probability density
\begin{equation}
\rho_{\phi}=i\left(\varphi^{*}\partial_{\phi}\varphi-\varphi\partial_{\phi}\varphi^{*}\right).
\label{12.14}
\end{equation}
Replacing the wave function \eqref{12.3} assuming $c_{1}=0,c_{2}=1$, we
obtain the non-normalized probability density 
\begin{equation}
\rho_{\phi}=-\cfrac{1}{k_{\alpha}\phi},
\label{12.15}
\end{equation}
which is positive if $k_{\alpha}<0$. Considering that there are no
monochromatic physical wave functions, we construct the Gaussian wave
packet which is a solution of the Wheeler-De Witt equation \eqref{12.2}
\begin{equation}
\Phi\left(\alpha,\phi\right)=\int_{-\infty}^{\,0}dk_{\alpha}A(k_{\alpha})e^{ik_{\alpha}\alpha}\left(\cfrac{1}{2ik_{\alpha}}\right)\phi^{2ik_{\alpha}},
\label{12.16}
\end{equation}
where 
\begin{equation}
A\left(k_{\alpha}\right)=\cfrac{1}{\sqrt{\left(2\pi\right)}\sigma_{\alpha}}\;e^{-\left(\frac{\left(k_{\alpha}-\bar{k}_{\alpha}\right)^{2}}{2\sigma_{\alpha}^{2}}\right)}.
\label{12.17}
\end{equation}

The normalized probability density 
\begin{equation}
\rho_{\phi}=i\left(\Phi^{*}\partial_{\phi}\Phi-\Phi\partial_{\phi}\Phi^{*}\right).
\label{12.18}
\end{equation}
gives the probability of finding the FLRW Universe at a certain instant
$\phi$ per unit of ``spatial coordinate'' $\alpha$.

\begin{figure}[h]
\includegraphics[scale=0.3]{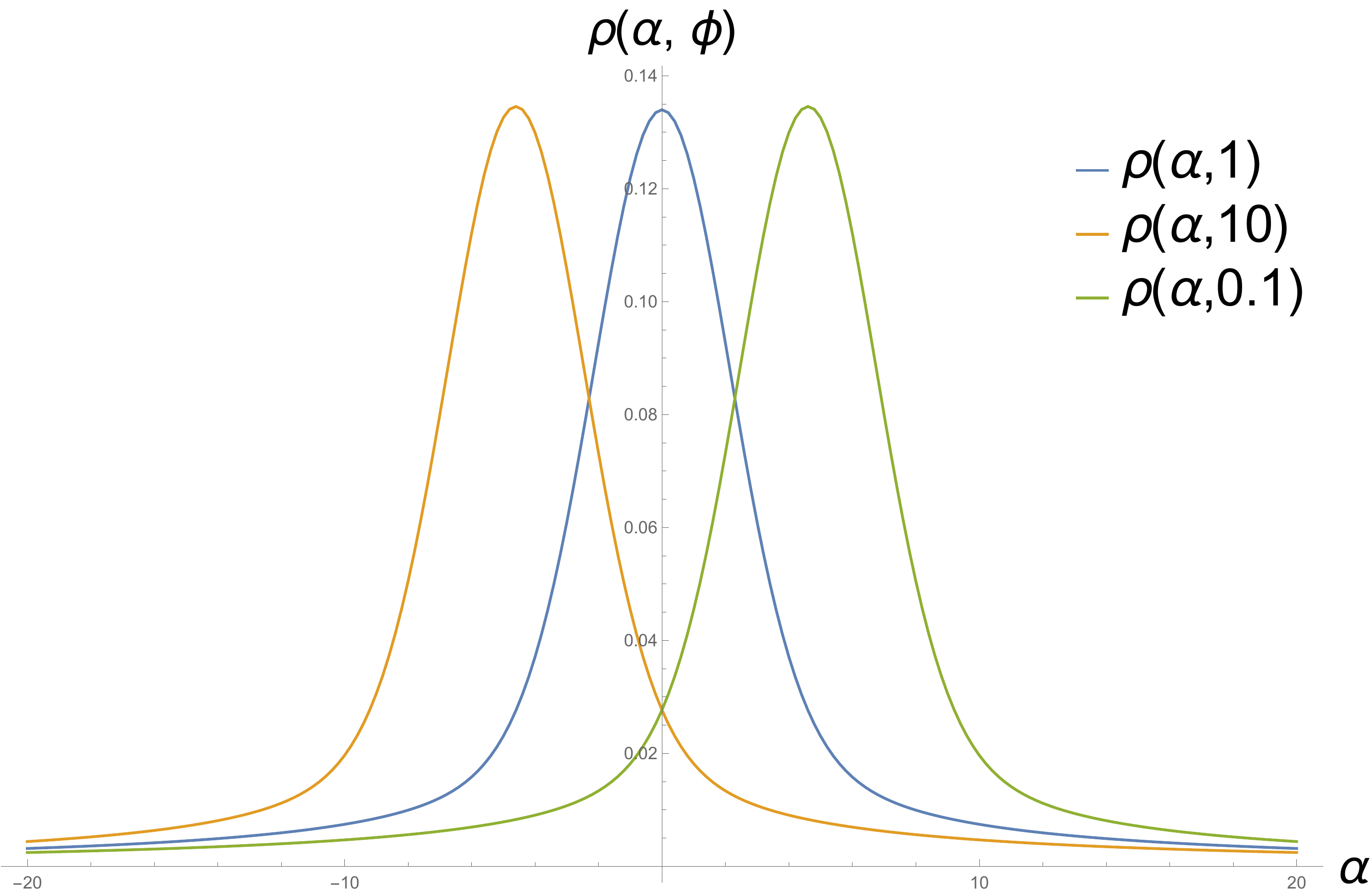}
\caption{\label{fig5}The trend of the probability density $\rho\left(\alpha,\phi\right)$
in function of the coordinate $\alpha$ for different values of $\phi$.
$\phi=1$ is shown in blue; $\phi=\frac{1}{10}$ is shown in green;
$\phi=1$0 is shown in orange.}
\end{figure}

We now study
the trend of the quantity \eqref{12.18} with respect to   $\phi>0$  as we can see from \eqref{12.13}
(FIG.\ref{fig5}).

\begin{figure}[h]
\begin{centering}
\includegraphics[scale=0.3]{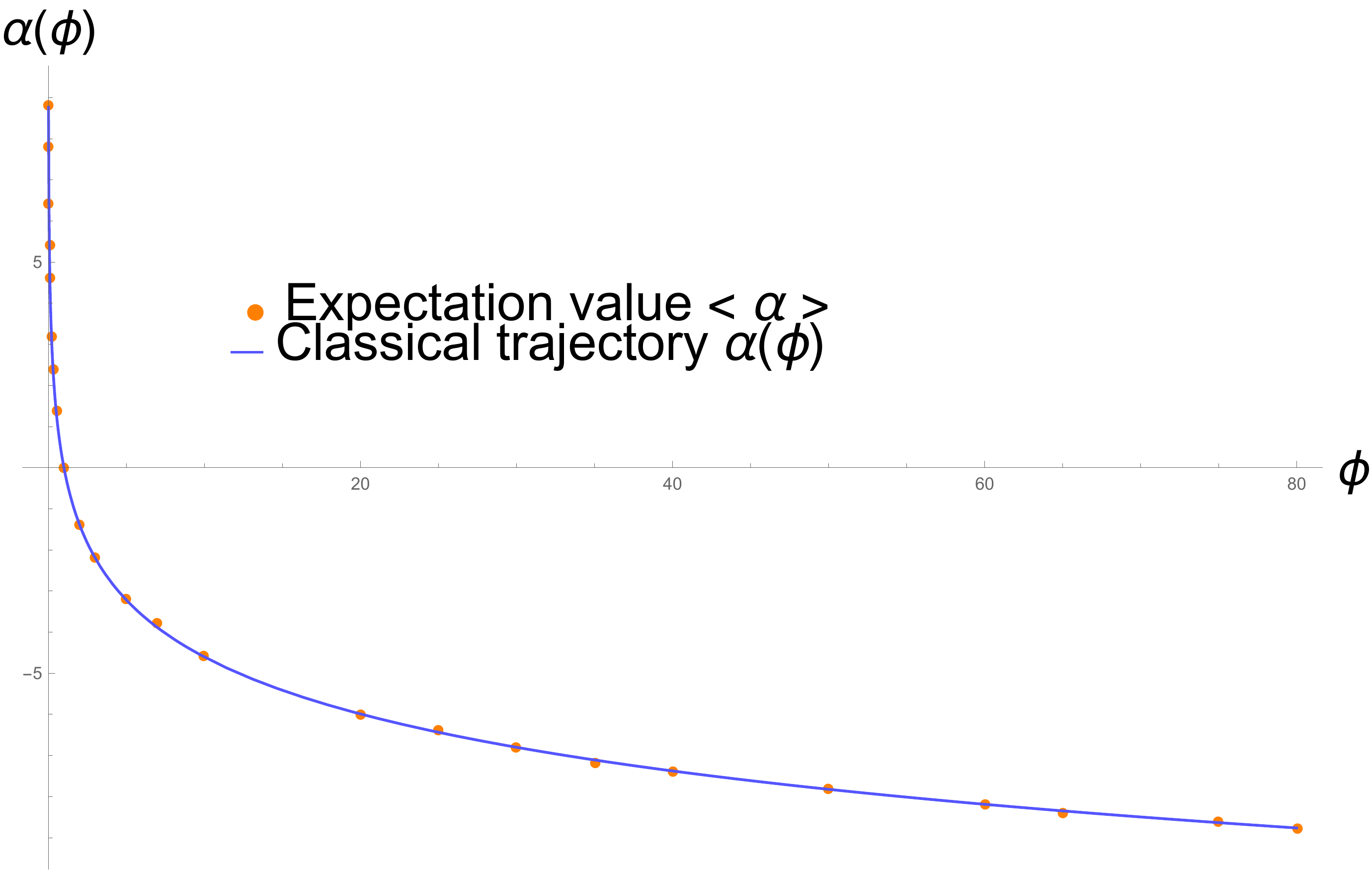} 
\par\end{centering}
\caption{\label{fig6}Comparison between the classical trajectory (solid
blue line) and the values $\alpha\left(\phi\right)$ where the probability
density is maximally localized as $\phi$ varies (dotted orange line).}
\end{figure}

From FIG.\ref{fig5} we see that there is no spreading of the wave packet
and the probability density remains perfectly localized over time.
We now prove the validity of the Ehrenfest Theorem by comparing the
classical trajectory provided by Hamilton's equations and the
values of the coordinates $\alpha$ corresponding to the maximums
of the probability density $\rho_{\phi}$ as $\phi$ varies. Strictly speaking, we should study the expectation value
$<\Phi|\hat{\alpha}|\Phi>$, but for highly localized Gaussian probability
density the values of $\alpha$ where there's a maximum is a good
approximation). From FIG.\ref{fig6} it is evident that the quantum
dynamics perfectly follows the classical dynamics of the FLRW Universe
in the Jordan frame. This result demonstrates that such a quantum-cosmological
model reaches the singularity ($\phi\rightarrow+\infty$) in a classical
way, allowing us to consider the quantum effects in the Planckian
regime, as effects of lower order on a semiclassical Universe.

\subsection{Comparison with the FLRW model filled with a scalar field in General
Relativity}

\begin{figure}[h]
\begin{centering}
\includegraphics[scale=0.3]{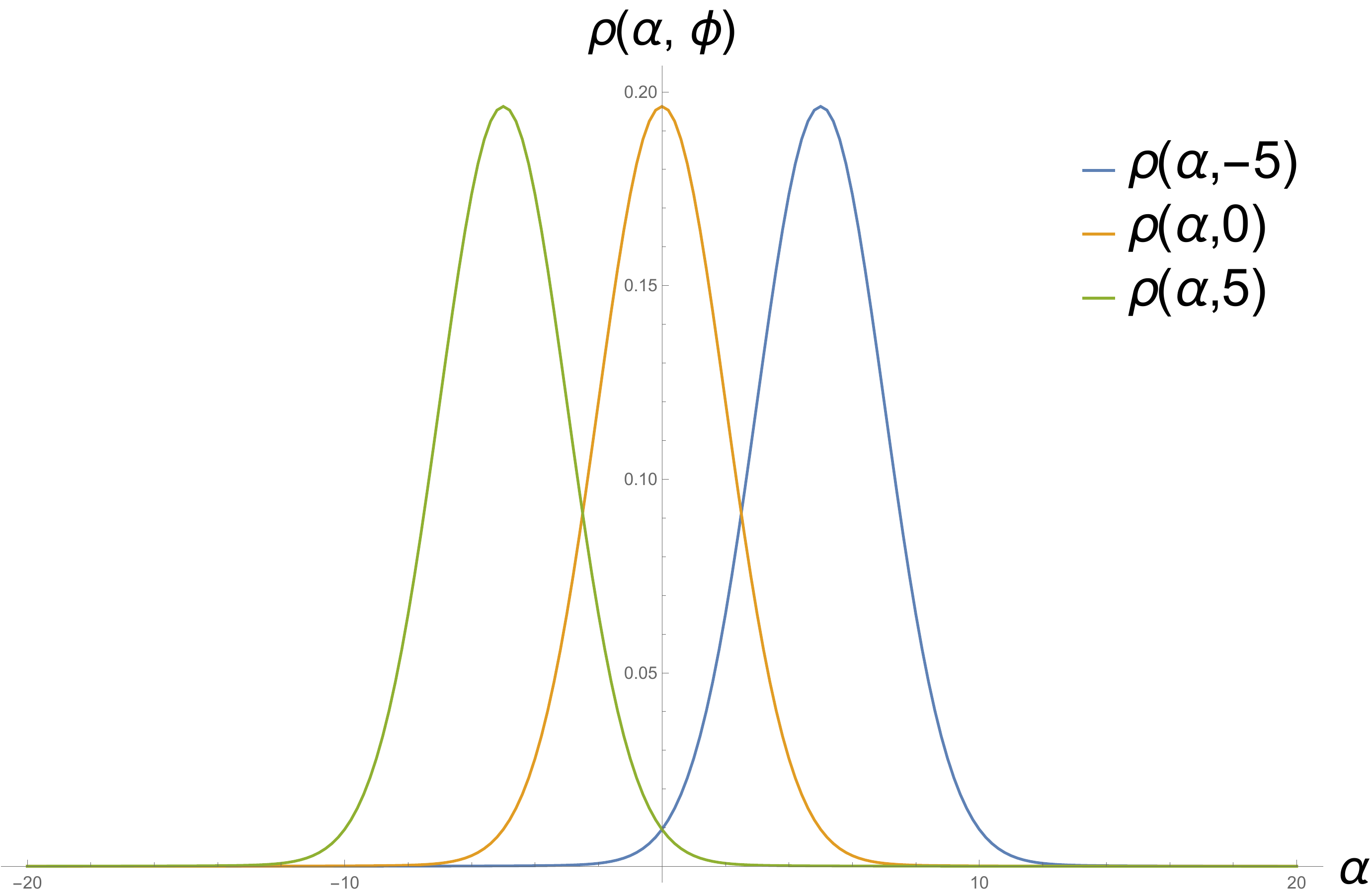} 
\par\end{centering}
\caption{\label{fig7}Evolution of the probability density $\rho\left(\alpha,\phi\right)$
in function of the coordinate $\alpha$ for different values of $\phi$.
$\phi=0$ is shown in orange; $\phi=5$ is shown in green; $\phi=-5$
is shown in blue.}
\end{figure}

The Wheeler-De Witt equation in Misner variables for the FLRW Universe
in GR filled with a minimally coupled scalar field $\phi$ is 
\begin{equation}
\begin{array}{ccl}
\mathcal{H}_{FLRW+\phi} & = & \cfrac{2 e^{-\frac{3}{2}\alpha}}{\mathbf{v}}\left[-\hat{p}_{\alpha}^{2}+\hat{\pi}_{\phi}^{2}\right]|\varphi\rangle\\
 & = & \cfrac{2 e^{-\frac{3}{2}\alpha}}{\mathbf{v}}\left(\partial_{\alpha}^{2}-\partial_{\phi}^{2}\right)\varphi\left(\alpha,\phi\right)=0,
\end{array}
\label{12.19}
\end{equation}
where we neglected the potential term $V\left(\phi\right)$ related
to the derivatives of the scalar field. Differently from the case
\eqref{12.2} in the Jordan frame, this equation resembles a 2-dimensional
Klein-Gordon equation describing a free and massless particle, where
the external scalar field $\phi$ plays the role of the time variable.
The wave function $\varphi$ solution of the equation is a plane wave
\begin{equation}
\varphi\left(\alpha,\phi\right)=Ae^{i\left(k_{\alpha}\alpha+k_{\phi}\phi\right)},
\label{12.20}
\end{equation}
and assuming a Gaussian wave packet at the initial time $\phi_{0}$,
we write the general solution 
\begin{equation}
\Phi\left(\alpha,\phi\right)=\int_{-\infty}^{\,0}dkA(k)e^{ik\left(\alpha+\phi\right)}
\label{12.21}
\end{equation}
where 
\begin{equation}
A\left(k\right)=\cfrac{1}{\sqrt{\left(2\pi\right)}\sigma}\;e^{-\left(\frac{\left(k-\overline{k}\right)^{2}}{2\sigma^{2}}\right)},
\label{12.22}
\end{equation}
and $k\equiv k_{\phi}=k_{\alpha}<0$. Through the study of the probability
density 
\begin{equation}
\rho_{\phi}=i\left(\Phi^{*}\partial_{\phi}\Phi-\Phi\partial_{\phi}\Phi^{*}\right),
\label{12.23}
\end{equation}
we can infer the absence of the spreading phenomenon typical of a
free quantum particle (FIG.\ref{fig7}).

We conclude that even in the case of GR with a minimally coupled scalar
field, the FLRW Universe admits a classical limit valid up to Planckian
regimes. In support of this claim, we apply the ADM reduction procedure
and derive the Hamilton's equations 
\begin{equation}
\left\{ \begin{array}{l}
\cfrac{\partial\alpha}{\partial\phi}=\cfrac{\partial h_{\phi}}{\partial p_{\alpha}}=-\cfrac{1}{2\sqrt{p_{\alpha}}},\\
\cfrac{\partial p_{\alpha}}{\partial\phi}=-\cfrac{\partial h_{\phi}}{\partial\alpha}=0,
\end{array}\right.
\label{12.23}
\end{equation}
from which the classical trajectory takes the form 
\begin{equation}
\alpha\left(\phi\right)=-c\phi,
\label{12.24}
\end{equation}
where $c$ is a constant of the motion. 
\begin{figure}
\begin{centering}
\includegraphics[scale=0.3]{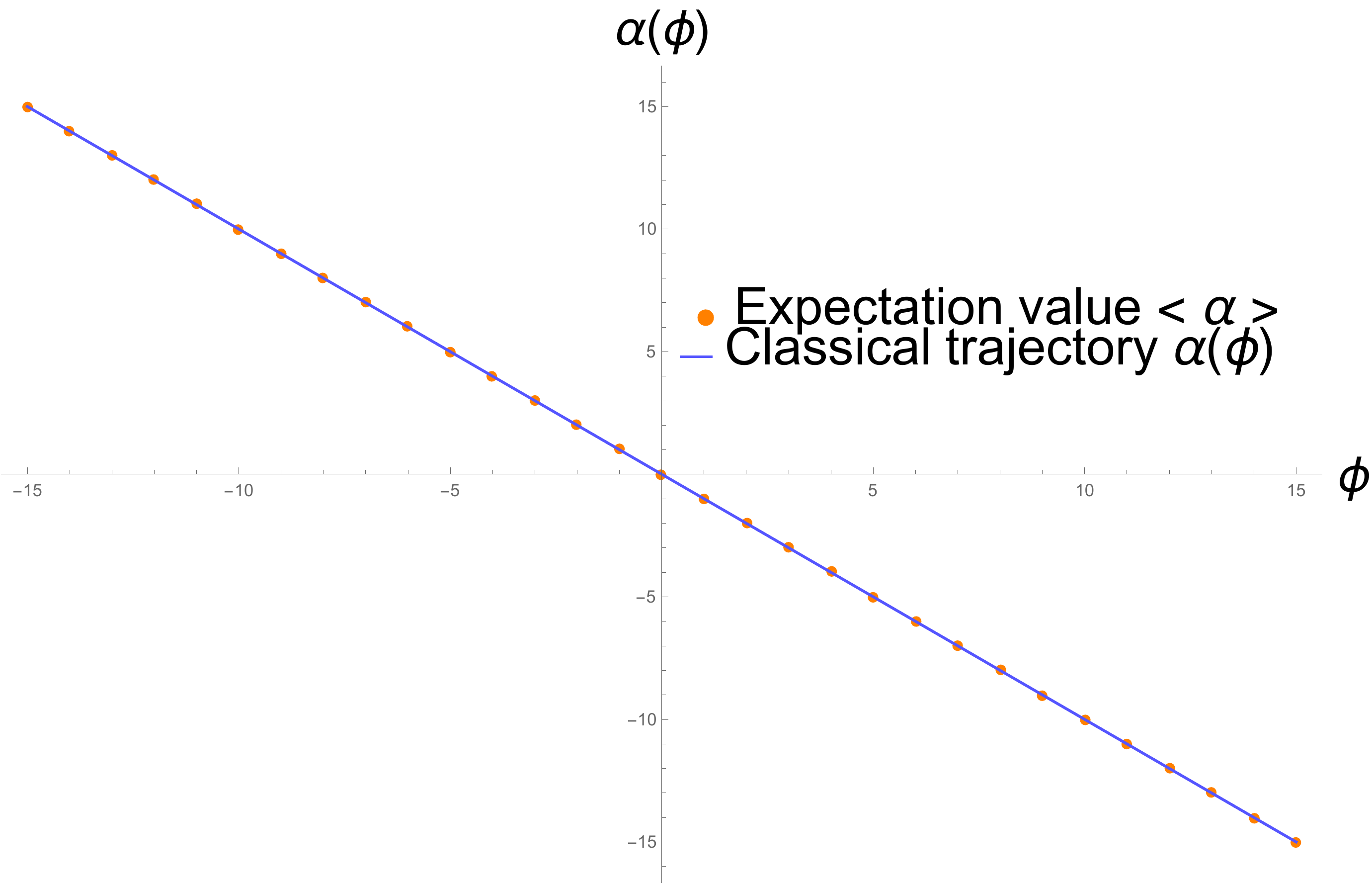} 
\par\end{centering}
\caption{\label{fig8}Comparison between the classical trajectory (solid
blue line) and the values $\alpha\left(\phi\right)$ where the probability
density is maximally localized as $\phi$ varies (dotted orange line).}
\end{figure}

From FIG.\ref{fig8} it emerges that quantum dynamics perfectly follows
the classical dynamics of the FLRW Universe filled with a minimally
coupled scalar field. Furthermore, it is possible to fix the constant
of motion $c=1$. In conclusion, we stress two main differences between
the two models under consideration: the first one is the linear vs.
logarithmic trend of $\alpha\left(\phi\right)$ with which the singularity
is approached; the second one is the range of values assumed by the
time variable $\phi$ which, in the case of the Jordan frame, was
limited only to the positive semi-axis of the Real numbers. In both
models, however, the canonical quantization based on the Wheeler-De
Witt formalism does not solve the problem of the existence of a cosmological
singularity, paving the way for different theoretical approaches.

\part{Concluding remarks}

In this paper, we provided a systematic and detailed analysis of the
cosmological implication of an $f(R)$-gravity in the Jordan frame,
especially in the limit of a negligible potential term of the non-minimally
coupled scalar field. Some interesting results have been clearly stated
and here briefly re-analyzed. As first step, we show in the general
Hamiltonian formulation, as well as in its implementation to the dynamics
of a generic universe, that if we use the 3-metric determinant as
a separated variable, like in \cite{PhysRev.160.1113}, its
conjugate momentum enters linearly only in the scalar Hamiltonian
constraint. Differently from what it takes place in the standard Wheeler-DeWitt
equation, we are here always able to set up a Schröedinger formulation
in which the 3-determinant plays the role of a time variable, although
the associated Hamiltonian turns out to be a non-local functional
operator. The second very significant result we developed consists
of the proof that the Bianchi IX cosmology is no longer characterized
by a chaotic dynamics in an extended $f(R)$ theory of gravity, as
soon as the potential term bringing the information on the form of
$f$ is negligible near to the singularity. Here we adopted Minser variables
to describe the Bianchi IX line element and we demonstrated both the
existence and the actractivity of a Kasner stability region. The definition
of the region where a Kasner regime becomes stable is derived analytically,
while the proof that the system always reaches such a configuration has
been performed on a numerical level. This latter study relies on the
existence of three constants of the motion and it can be regarded
as the natural generalization of the original Misner approach \cite{PhysRev.186.1319}.
The result obtained here about the chaos suppression is coherent with
the analysis in \cite{PhysRevD.90.101503},
where the same issue is discussed using a $f(R)$ approach in the
so-called Einstein framework and adopting Misner-Chitré-like variables.
Finally, we touch the question of the canonical quantization of the
Bianchi I dynamics, concentrating our attention to the quantum evolution
of an isotropic Universe. For the Bianchi I Wheeler-DeWitt equation a conserved current
is constructed and a probability density is identified
by interpreting the non-minimal scalar field as the internal time
variable. A localized wave packet is then constructed for the isotropic
universe and we show how the singularity is still present in such
a quantum scenario. Furthermore a comparison between the evolutionary
cosmologies using as internal time a minimally and a non-minimally
coupled scalar field respectively is provided. Apart from a different
dynamics of the peak of the localized packet, we clarify how the non-minimally
coupled scalar field emerging in a Jordan frame is a valuable internal
time and opens a new point of view on the interpretation of modified
$f(R)$ theory of gravity: the additional scalar mode, summarizing
the functional form $f$, can be interpreted as a time-like degree
of freedom for the gravitational quantum dynamics, although a viable
approach could also emerge by dealing, as stressed above, with the
3-metric determinant. Since the Bianchi model dynamics has paradigmatic
features of the generic inhomogeneous cosmology, we are lead to infer
that the present analysis has a relevant impact on a more general
sector of the cosmological problem. However, it must be recalled that,
when the potential term associated to the non-minimally coupled scalar
field is no longer negligible near the singularity, its presence can
significantly affect the validity of some of the present issues, including
the non-chaoticity of the Bianchi IX Universe.

\section{Appendix A}

We explicit the canonical transformation used in Part IV, sec.VI
\begin{equation}
\begin{array}{l}
p^{ij}\partial_{t}h_{ij}\\
=\underset{a}{\sum}p_{a}\partial_{t}q_{a}\,+\,\underset{a}{\sum}\pi_{a}\partial_{t}y_{a}\\
p^{ij}\partial_{t}\left(\underset{a}{\sum}e^{q_{a}}e_{i}^{a}e_{j}^{a}\right)\\
=\underset{a}{\sum}p_{a}\partial_{t}q_{a}\,+\,\underset{a}{\sum}\pi_{a}\partial_{t}y_{a}\\
p^{ij}\underset{a}{\sum}\partial_{t}q_{a}e^{q_{a}}e_{i}^{a}e_{j}^{a}+2p^{ij}\underset{a}{\sum}e^{q_{a}}e_{i}^{a}\partial_{t}e_{j}^{a}\\
=\underset{a}{\sum}p_{a}\partial_{t}q_{a}\,+\,\underset{a}{\sum}\pi_{a}\partial_{t}y_{a}\\
p^{ij}\underset{a}{\sum}\partial_{t}q_{a}e^{q_{a}}e_{i}^{a}e_{j}^{a}+2p^{ij}\underset{a}{\sum}e^{q_{a}}e_{i}^{a}\partial_{t}\left(O_{c}^{a}\partial_{j}y^{c}\right)\\
=\underset{a}{\sum}p_{a}\partial_{t}q_{a}\,+\,\underset{a}{\sum}\pi_{a}\partial_{t}y_{a}\\
p^{ij}\underset{a}{\sum}\partial_{t}q_{a}e^{q_{a}}e_{i}^{a}e_{j}^{a}+\partial_{j}\left(2p^{ij}\underset{a}{\sum}e^{q_{a}}e_{i}^{a}\partial_{t}y^{c}O_{c}^{a}\right)\\
-\partial_{j}\left(2p^{ij}\underset{a}{\sum}e^{q_{a}}e_{i}^{a}O_{c}^{a}\right)\partial_{t}y^{c}\\
=\underset{a}{\sum}p_{a}\partial_{t}q_{a}\,+\,\underset{a}{\sum}\pi_{a}\partial_{t}y_{a}\\
p^{ij}\underset{a}{\sum}\partial_{t}q_{a}e^{q_{a}}e_{i}^{a}e_{j}^{a}-\underset{a}{\sum}\partial_{j}\left(2p^{ij}e^{q_{a}}e_{i}^{a}O_{c}^{a}\right)\partial_{t}y^{c}\\
=\underset{a}{\sum}p_{a}\partial_{t}q_{a}\,+\,\underset{a}{\sum}\pi_{a}\partial_{t}y_{a}
\end{array}
\label{A.1}
\end{equation}
where we neglected the the term $\partial_{j}\left(2p^{ij}\underset{a}{\sum}e^{q_{a}}e_{i}^{a}\partial_{t}y^{c}O_{c}^{a}\right)$,
because $p^{ij}$ is a tensor density. From the definition of the
\emph{superHamiltonian} of gravity in the Jordan frame 
\begin{align}
\mathcal{H}_{g,JF} & =\cfrac{2}{\sqrt{h}}\left(\cfrac{p_{ij}p^{ij}-\frac{1}{3}p^{2}}{\phi}+\cfrac{1}{6}\phi\pi_{\phi}^{2}-\cfrac{1}{3}p\pi_{\phi}\right)\nonumber \\
 & \quad+\cfrac{\sqrt{h}}{2}\left(V(\phi)-\phi\,^{3}R+2D_{i}D^{i}\phi\right)
\end{align}
we replace the relationships 
\begin{align}
p^{ij}p_{ij} & =\underset{a}{\sum}e^{-q_{a}}p_{a}e_{a}^{i}e_{a}^{j}\underset{b}{\sum}e^{q_{b}}p_{b}e_{i}^{b}e_{j}^{b}\nonumber \\
 & =\underset{a}{\sum}\underset{b}{\sum}e^{-q_{a}}e^{q_{b}}p_{a}p_{b}\delta_{a}^{b}e_{a}^{j}e_{j}^{b}\nonumber \\
 & =\underset{a}{\sum}p_{a}^{2}
\end{align}
\begin{align}
p^{2} & =\left(h_{ij}p^{ij}\right)^{2}\nonumber \\
 & =\left(\underset{a}{\sum}e^{q_{a}}e_{i}^{a}e_{j}^{a}\underset{b}{\sum}e^{-q_{b}}p_{b}e_{b}^{i}e_{b}^{j}\right)^{2}\nonumber \\
 & =\left(\underset{a}{\sum}\underset{b}{\sum}e^{q_{a}}e^{-q_{b}}p_{b}\delta_{b}^{a}e_{j}^{a}e_{b}^{j}\right)^{2}\nonumber \\
 & =\left(\underset{a}{\sum}p_{a}\right)^{2}
\end{align}
in the kinetic term of the \emph{superHamiltonian }obtaining the \emph{superHamiltonian} \eqref{6.6}
 
\begin{align}
\mathscr{H}_{g,JF} & =\cfrac{2}{\sqrt{h}}\left(\cfrac{\underset{a}{\sum}p_{a}^{2}-\frac{1}{3}\left(\underset{a}{\sum}p_{a}\right)^{2}}{\phi}+\cfrac{1}{6}\phi\pi_{\phi}^{2}-\cfrac{1}{3}\underset{a}{\sum}p_{a}\pi_{\phi}\right)\nonumber \\
 & \quad+\cfrac{\sqrt{h}}{2}\left(V(\phi)-\phi\,^{3}R+2D_{i}D^{i}\phi\right)
\end{align}
In order to obtain the \emph{supermomentum} \eqref{6.7}, we redefine the
\emph{supermomentum} of gravity in the Jordan frame 
\begin{equation}
\mathcal{H}_{i}^{g,JF}=-2D_{j}p_{i}^{j}+\pi_{\phi}\partial_{i}\phi
\end{equation}
and rewriting $p_{i}^{j}=\sqrt{h}T_{i}^{j}$ we get 
\begin{align}
\mathcal{H}_{i}^{g} & -\pi_{\phi}\partial_{i}\phi\nonumber \\
 & =-2\left(\partial_{j}p_{i}^{j}-\Gamma_{ij}^{l}p_{l}^{j}\right)\nonumber \\
 & =-2\left[\partial_{j}\left(\underset{a}{\sum}p_{a}e_{i}^{a}e_{a}^{j}\right)-\Gamma_{ij}^{l}p_{l}^{j}\right]\nonumber \\
 & =-2\left[\partial_{j}\left(\underset{a}{\sum}p_{a}e_{i}^{a}e_{a}^{j}\right)-\cfrac{1}{2}\left(\partial_{i}h_{jl}+\partial_{j}h_{il}-\partial_{l}h_{ij}\right)p^{jl}\right]\nonumber \\
 & =-2\left[\partial_{j}\left(\underset{a}{\sum}p_{a}e_{i}^{a}e_{a}^{j}\right)-\cfrac{1}{2}\left(\partial_{i}h_{jl}\right)p^{jl}\right]\nonumber \\
 & =-2\underset{a}{\sum}\left(\partial_{j}p_{a}\delta_{j}^{i}+p_{a}\partial_{j}\left(e_{a}^{j}O_{b}^{a}\partial_{i}y^{b}\right)\right)\nonumber \\
 & \quad+\partial_{i}\left(\underset{d}{\sum}e^{q_{d}}e_{j}^{d}e_{l}^{d}\right)\underset{c}{\sum}e^{-q_{c}}p_{c}e_{c}^{j}e_{c}^{l}\nonumber \\
 & =-2\underset{a}{\sum}\left(\partial_{i}p_{a}+p_{a}\partial_{j}\left(e_{a}^{j}O_{b}^{a}\right)\partial_{i}y^{b}+p_{a}e_{a}^{j}O_{b}^{a}\partial_{j}\partial_{i}y^{b}\right)\nonumber \\
 & \quad+\underset{c}{\sum}\left(\partial_{i}q_{c}e^{q_{c}}e_{j}^{c}e_{l}^{c}e^{-q_{c}}p_{c}e_{c}^{j}e_{c}^{l}\right)+\nonumber \\
 & \quad+\underset{c}{\sum}\left(e^{q_{c}}\partial_{i}\left(e_{j}^{c}\right)e_{l}^{c}e^{-q_{c}}p_{c}e_{c}^{j}e_{c}^{l}\right)+\underset{c}{\sum}e^{q_{c}}e_{j}^{d}\partial_{i}\left(e_{l}^{c}\right)e^{-q_{c}}p_{c}e_{c}^{j}e_{c}^{l}\nonumber \\
 & =-2\underset{a}{\sum}\left(\partial_{i}p_{a}+p_{a}\partial_{j}\left(e_{a}^{j}O_{b}^{a}\right)\partial_{i}y^{b}+p_{a}e_{a}^{j}O_{b}^{a}\partial_{j}\partial_{i}y^{b}\right)\nonumber \\
 & +\underset{c}{\sum}\left(p_{c}\partial_{i}q_{c}+2p_{c}\partial_{i}\left(O_{d}^{c}\partial_{l}y^{d}\right)e_{c}^{l}\right)\nonumber \\
 & =-2\underset{a}{\sum}\left(\partial_{i}p_{a}+p_{a}\partial_{j}\left(e_{a}^{j}O_{b}^{a}\right)\partial_{i}y^{b}+p_{a}e_{a}^{j}O_{b}^{a}\partial_{j}\partial_{i}y^{b}\right)\nonumber \\
 & \quad+\underset{c}{\sum}p_{c}\partial_{i}q_{c}++2\underset{c}{\sum}p_{c}\partial_{i}O_{d}^{c}\left(\partial_{l}y^{d}e_{c}^{l}\right)\nonumber \\
 & \quad+2\underset{c}{\sum}p_{c}O_{d}^{c}\partial_{i}\partial_{l}y^{d}e_{c}^{l}\nonumber \\
 & =-2\underset{a}{\sum}\left(\partial_{i}p_{a}+p_{a}\partial_{j}\left(e_{a}^{j}O_{b}^{a}\right)\partial_{i}y^{b}\right)\nonumber \\
 & \quad-2\underset{a}{\sum}\left(p_{a}e_{a}^{j}O_{b}^{a}\partial_{j}\partial_{i}y^{b}+p_{a}e_{a}^{l}O_{d}^{a}\partial_{i}\partial_{l}y^{d}\right)+\nonumber \\
 & \quad+\underset{c}{\sum}p_{c}\partial_{i}q_{c}+2\underset{c}{\sum}p_{c}\partial_{i}O_{d}^{c}\left(O^{-1}\right)_{c}^{d}\nonumber \\
 & =-2\underset{a}{\sum}\left(\partial_{i}p_{a}+p_{a}\partial_{j}\left(e_{a}^{j}O_{b}^{a}\right)\partial_{i}y^{b}\right)\nonumber \\
 & \quad+\underset{c}{\sum}p_{c}\partial_{i}q_{c}+2\underset{c}{\sum}p_{c}\partial_{i}O_{d}^{c}\left(O^{-1}\right)_{c}^{d}
\end{align}
From \eqref{A.1} we define 
\begin{align}
\pi_{c} & =-2\underset{a}{\sum}\partial_{j}\left(p_{a}e_{a}^{j}(t,x^{k})O_{c}^{a}(x^{k})\right)\nonumber \\
 & =-2\underset{a}{\sum}\left(\partial_{j}p_{a}O_{c}^{a}O_{a}^{d}\cfrac{\partial x^{j}}{\partial y^{d}}+p_{a}\partial_{j}O_{c}^{a}e_{a}^{j}+p_{a}O_{c}^{a}\partial_{j}e_{a}^{j}\right)\nonumber \\
 & =-2\underset{a}{\sum}\left(\partial_{j}p_{a}\cfrac{\partial x^{j}}{\partial y^{c}}+p_{a}\partial_{j}O_{c}^{a}e_{a}^{j}+p_{a}O_{c}^{a}\partial_{j}e_{a}^{j}\right)
\end{align}
from which 
\begin{align}
\pi_{c}\partial_{i}y^{c} & =-2\underset{a}{\sum}\left(\partial_{j}p_{a}\cfrac{\partial x^{j}}{\partial y^{c}}+p_{a}\partial_{j}O_{c}^{a}e_{a}^{j}+p_{a}O_{c}^{a}\partial_{j}e_{a}^{j}\right)\cfrac{\partial y^{c}}{\partial x^{i}}\nonumber \\
 & =-2\underset{a}{\sum}\left(\partial_{i}p_{a}+p_{a}\partial_{j}O_{c}^{a}e_{a}^{j}\partial_{i}y^{c}+p_{a}O_{c}^{a}\partial_{j}e_{a}^{j}\partial_{i}y^{c}\right)\nonumber \\
 & =-2\underset{a}{\sum}\left(\partial_{i}p_{a}+p_{a}\partial_{j}\left(e_{a}^{j}O_{c}^{a}\right)\partial_{i}y^{c}\right)
\end{align}
We, finally, find the \emph{supermomentum }in the general framework \eqref{6.7} 
\begin{equation}
\mathscr{H}_{i,JF}^{B}=\pi_{\phi}\partial_{i}\phi+\pi_{a}\partial_{i}y^{a}+p_{a}\partial_{i}q^{a}+2p_{a}\left(O^{-1}\right)_{a}^{b}\partial_{i}O_{b}^{a}
\end{equation}

\section{Appendix B}

As we have seen in section X, we have defined a stability region within
which we are sure that the particle-Universe doesn't impact with the
potential wall, scattering from a Kasner solution to another. The
calculation of the boundaries of the \emph{Kasner stability region}
was performed with the help of \emph{Wolfram Mathematica}.

Now we want to describe the complete solutions:

\begin{equation}
\Biggl\{\begin{array}{l}
-1<\beta_{+}^{'}<0\\
\beta_{-}^{'}>Root\left[Ax^{3}+Bx^{2}+Cx+D\right] \\
\beta_{-}^{'}<Root\left[Ax^{3}-Bx^{2}+Cx-D\right]
\end{array}
\end{equation}

\begin{equation}
\Biggl\{\begin{array}{l}
\beta_{+}^{'}=0\\
\beta_{-}^{'}>Root\left[1+3x^{2}+4\sqrt{3}x^{3}\right]\\
\beta_{-}^{'}<Root\left[-1-3x^{2}+4\sqrt{3}x^{3}\right]
\end{array}
\end{equation}

\begin{equation}
\Biggl\{\begin{array}{l}
0<\beta_{+}^{'}<\frac{1}{2}\\
\beta_{-}^{'}>Root\left[Ax^{3}+Bx^{2}+Cx+D\right]\\
\beta_{-}^{'}<Root\left[Ax^{3}-Bx^{2}+Cx-D\right]
\end{array}
\end{equation}

\begin{equation}
\Biggl\{\begin{array}{l}
\frac{1}{2}<\beta_{+}^{'}<Root\left[-1-3x^{2}+8x^{3}\right]\\
-\sqrt{\frac{1+3v_{+}^{2}-8v_{+}^{3}}{-3+8v_{+}}}<\beta_{-}^{'}<\sqrt{\frac{1+3v_{+}^{2}-8v_{+}^{3}}{-3+8v_{+}}}
\end{array}
\end{equation}
with: $A=4\sqrt{3},B=3+4v_{+},C=4\sqrt{3}$ and $D=1+3v_{+}^{2}+4v_{+}^{3}$.

\newpage

\bibliography{artbib}
\end{document}